\documentclass[11pt,letterpaper,notoc]{JHEP3}

  \usepackage{latexsym,bm,amsmath,amssymb,amsfonts}
  \usepackage{epsfig,graphics,graphicx,subfigure}
  \usepackage{cite}
  \usepackage{slashed}

  \long\def\comment#1{ }
  \newcommand{\dif}{{\rm d}}
  \newcommand{\dY}{\dif Y}

  \newcommand{\del}{\partial}
  \newcommand{\lan}{\langle}
  \newcommand{\ran}{\rangle}
  \newcommand{\mcal}{\mathcal}
  \newcommand{\rme}{{\rm e}}
  
  \newcommand{\rmL}{{\rm L}}
  \newcommand{\rmR}{{\rm R}}

  \newcommand{\Lam}{\Lambda_{{\rm QCD}}}
  
  \newcommand{\order}[1]{\mcal{O}{(#1)}}
  \newcommand{\nn}{\nonumber\\}

  \newcommand{\beq}{\begin{eqnarray}}
  \newcommand{\eeq}{\end{eqnarray}}
  \newcommand{\avg}[1]{\langle #1 \rangle}
  \def\simge{\mathrel{%
   \rlap{\raise 0.511ex \hbox{$>$}}{\lower 0.511ex \hbox{$\sim$}}}}
  \def\simle{\mathrel{
   \rlap{\raise 0.511ex \hbox{$<$}}{\lower 0.511ex \hbox{$\sim$}}}}

\title{Pomeron loop and running coupling effects in high energy QCD evolution}

\author{A.~Dumitru \\Institut f\"ur Theoretische Physik,
J.~W.~Goethe Universit\"at, Max-von-Laue Strasse 1,
D-60438 Frankfurt am Main, Germany\\ E-mail: \email{dumitru@th.physik.uni-frankfurt.de}}
\author{E.~Iancu\thanks{Membre du Centre National de la Recherche Scientifique
(CNRS), France.},\, L.~Portugal\thanks{On leave from Instituto de F\'isica,
Universidade Federal do Rio de Janeiro, Brazil}
\\ Service de Physique Th\'eorique de Saclay,
CEA/DSM/SPhT, F-1191 Gif-sur-Yvette, France\\ E-mail: \email{edmond.iancu@cea.fr},
\email{licinio@if.ufrj.br}}
\author{G.~Soyez\thanks{On leave from the Fundamental Theoretical
Physics group of the University of Li\`ege,
Belgium}\\Physics Department, Brookhaven National Laboratory,
Upton, NY 11973, USA\\ E-mail: \email{G.Soyez@ulg.ac.be}}
\author{D.N.~Triantafyllopoulos\\
ECT$^*$, Villa Tambosi, Strada delle Tabarelle 286, I-38050 Villazzano (TN),
Italy\\ E-mail: \email{dionysis@ect.it}}

\abstract{Within the framework of a (1+1)--dimensional model which mimics
evolution and scattering in QCD at high energy, we study the influence of the
running of the coupling on the high--energy dynamics with Pomeron loops.
We find that the particle number fluctuations are strongly suppressed by
the running of the coupling, by at least one order of magnitude as compared
to the case of a fixed coupling, for all the rapidities that we have
investigated, up to $Y=200$. This reflects the slowing down of the evolution
by running coupling effects, in particular, the large rapidity evolution
which is required for the formation of the saturation front via diffusion.
We conclude that, for all energies of interest, processes like
deep inelastic scattering or forward particle production can be reliably
studied within the framework of a mean--field approximation (like the
Balitsky--Kovchegov equation) which includes running coupling effects.}

\begin{document}

\section{Introduction}\label{sect-intro}
\setcounter{equation}{0}

In the recent years, one has assisted at two major directions of progress
in our understanding of the QCD dynamics at high energy, both aiming at
improving over the previous status of the theory, as encoded in the
Balitsky--Kovchegov \cite{B,K} (BK) and
Jalilian-Marian--Iancu--McLerran--Weigert--Leonidov--Kovner (JIMWLK)
equations \cite{JKLW,W,CGC}. One of these directions started with the
recognition \cite{MS04,IMM04,IT04} of the fundamental role played by
gluon--number fluctuations (or `Pomeron loops') in the QCD evolution with
increasing energy, which led to more complete evolution equations (the
`Pomeron loop equations') \cite{IT04,IT05,MSW05,LL05,KL2005,BREM,Balit05}
and to a surprising link to problems in statistical physics \cite{IMM04},
which is rich in consequences \cite{IT04,BDMM,MPS05,HIMST06,GLUON,KSX06}.
The other line of research concentrated on the inclusion of
next--to--leading order (NLO) effects in the non--linear BK equation
\cite{Motyka,SCALING,MT02,DT02,MP032,RW03,IIT04,Nestor04}, and very
recently culminated in an explicit calculation of the running coupling
effects associated with fermion loops within perturbative QCD
\cite{GKRW,KWrun1,Brun,KWrun2,Alb07,FFP}. Both types of effects ---
Pomeron loops and NLO corrections --- turn out to be effects of
$\order{1}$, which modify in a dramatic way our previous expectations
based on the BK--JIMWLK equations. Yet, so far, these two types of
effects have never been considered together in a unified framework (in
particular, their mutual influence has never been addressed), mainly
because of the complexity of the Pomeron loop equations, which are quite
difficult to deal with already at leading--order.

In this paper, we shall for the first time study the effects of the
running of the coupling on the high--energy evolution with Pomeron loops,
within a one--dimensional model proposed in Ref. \cite{ISST07} which
captures in a simplified form the relevant dynamics in QCD. For this
model, that we shall extend here to the case of a running coupling (in
such a way to mimic the one--loop running coupling of QCD), we shall
provide a rather detailed numerical analysis of the `dilute--dense'
scattering process similar to deep inelastic scattering (DIS) in QCD. Our
results point out towards a very interesting, and rather unexpected,
conclusion: The `Pomeron loop' effects (i.e., the effects of
particle--number fluctuations) are {\em strongly suppressed} by the
running of the coupling --- by at least one order of magnitude as
compared to the case of a fixed coupling ---, in such a way that they
remain negligible up to the highest energy that we have investigated,
which corresponds to a rapidity $Y=200$. This finding has a corollary of
great practical interest: it implies that, for studies of the
high--energy dynamics (say, in the energy range at LHC) one can reliably
resort on appropriate {\it mean--field approximations}, so like the BK
equation properly generalized to include NLO (or, at least, running
coupling) effects.

Our numerical analysis is further corroborated by analytic estimates
which help clarifying the mechanism responsible for the suppression of
fluctuations: this is {\em not} the fact that, because of the running,
the value of the coupling is effectively smaller, as one might naively
think; rather, this is related to the fact that the running of the
coupling {\em drastically slows down} the evolution. With a running
coupling, not only the saturation momentum grows much slower with the
energy, but the same is also true for the ``BFKL diffusion'' \cite{BFKL},
which is the mechanism through which the saturation front (say, for the
gluon occupation number, or for the dipole scattering amplitude in DIS)
evolves towards its asymptotic, `geometric scaling', shape at high
energy. Namely, the diffusive radius grows with $Y$ like $Y^{1/6}$
(rather than the usual $Y^{1/2}$ behavior at fixed coupling), and such a
rise is too slow for the front to reach its asymptotic shape within the
energy range under consideration. Rather, the front preserves a
pre--asymptotic shape, which is not favorable for the growth of
fluctuations. Hence, during this whole `pre--asymptotic' evolution (that
we found to extend up to $Y=200$ at least), the DIS dynamics is the same
as predicted by the corresponding mean--field approximation (with running
coupling, of course).

Although obtained in a specific model, we are confident that these
conclusions should apply to QCD as well, for several reasons: First, as
just mentioned, these results are supported by analytic estimates which
are essentially identical in QCD and in the model under consideration.
Second, this model is truly similar to QCD: it has been constructed
\cite{BIT06,ISST07} by requiring boost--invariance together with an
evolution law inspired by the gluon evolution in the context of JIMWLK
equation (see Sect. \ref{sect-const} below for details). While simpler
than the original JIMWLK equation, because of the absence of color
degrees of freedom and the reduction to only one spatial dimension
(which, physically, plays the role of the gluon transverse momentum), the
model is at the same time more general, in that it allows for
particle--number fluctuations. In particular, the mean--field
approximation to this model turns out to be quite similar, even
quantitatively\footnote{In the sense of generating similar numerical
values for the anomalous dimension, the saturation exponent, etc.; see
Sect. \ref{sect-analytic}.}, to BK equation of QCD. Third, this model
belongs to the universality class of the statistical process known as
`reaction--diffusion' \cite{Saar,BD97}, so like the high--energy QCD
dynamics itself \cite{IMM04}. Although, strictly speaking, this
universality has been so far established only for the case of a fixed
coupling, we expect all the processes in this class to respond in a
similar way to the inclusion of a running coupling.

When applied to QCD, our conclusions could in particular explain why
phenomenological analyses inspired by the mean--field (BK) dynamics were
relatively successful in describing the small--$x$ experimental situation
at HERA, despite the fact that the proton wavefunction was expected to
develop strong correlations via Pomeron loops in the course of the
evolution : such correlations, which would transcend any mean--field
description, are in fact suppressed by the running of the coupling. Most
significantly, the phenomenon of {\em geometric scaling}
---  a hallmark of the BK dynamics \cite{SCALING,MT02,DT02,MP031,MP032} which
seems to be well verified by the HERA data \cite{geometric,MS06,GPSS06},
but which would be washed out by the fixed--coupling evolution with
fluctuations \cite{IMM04,IT04,HIMST06} --- is in fact resuscitated by our
present analysis. But this analysis also shows that, precisely due to
running coupling effects, the window for {\em strict} geometric scaling
is drastically reduced as compared to the respective mean--field estimate
at fixed coupling \cite{SCALING,MT02}. (The width of this window is
controlled by the `diffusive radius' alluded to above; see Sect.
\ref{sect-analytic} for details.) This conclusion is consistent with
previous studies of the BK equation with running coupling
\cite{MT02,DT02,MP032}, and also with the phenomenological analyses of
the HERA data \cite{BGBK,IIM03,GS07}, which need to include a substantial
amount of scaling violation in order to accurately describe the
experimental results.

This paper is organized as follows: In Section 2, we briefly review the
construction of the one--dimensional model \cite{ISST07} and show how
this can be accommodated with the running of the coupling. Section 3
offers a summary of known results concerning the dynamics with Pomeron
loops at fixed coupling, as well as the mean--field dynamics with running
coupling. Based on such previous results, we formulate our theoretical
expectations for the full problem (Pomeron loops and running coupling) at
the end of that section. These expectations are then confronted to
numerical results in Section 4, which is devoted to an extensive analysis
of the model. Section 5 contains our conclusions and some perspectives.

\section{Model description}\label{sect-const}
\setcounter{equation}{0}

The construction of the model from the underlying physical assumptions
--- boost invariance, multiple scattering in the eikonal approximation,
and evolution law {\em \`a la} JIMWLK (meaning that the probability for
emitting a new particle in one step of the evolution depends in a
non--linear way upon the density of particles created in the previous
steps, and saturates at high density)
--- has been described in detail in a previous publication
\cite{ISST07} (see also Refs. \cite{AMSalam95,BIT06} for earlier version
of the model, in zero spatial dimensions), so here we shall only describe
the generalization of this model to the case of a running coupling.

The ``hadronic systems'' which undergo evolution and scattering are two
systems of classical particles (`the right and left mover') distributed
in one spatial dimension (`the $x$--axis') which is transverse to the
collision axis. From the point of view of QCD, the position $x$ of the
particles along this transverse axis corresponds to the logarithm of the
transverse momentum of a gluon (or, within the QCD dipole picture
\cite{AM94}, to the logarithm of the inverse size of a dipole). The total
rapidity gap between the incoming systems is equal to $Y$, and we choose
the `laboratory frame' in such a way that the right (left) mover carries
a rapidity $Y-Y_0$ (respectively, $-Y_0$). The scattering is assumed to
be elastic, and the corresponding $S$--matrix $\avg{S}_Y$ to be real (as
typically the case in QCD at high energy). Needless to say, $\avg{S}_Y$
must be independent upon the rapidity divider $Y_0$, i.e., upon the
choice of the frame.

The particle configurations in the two systems are created via
high--energy evolution (from their respective rest--frames to the
laboratory frame), which is a stochastic process. Accordingly, these
configurations are themselves random, and can be described in the
language of probabilities: we shall denote by
$P_{\rmR}[n(x_{\rmR}),Y-Y_0]$ and, respectively,
$P_{\rmL}[m(x_{\rmL}),Y_0]$ the probability densities to find given
configurations in the two systems. Note that a configuration is described
as a function $n(x)$ (the density of particles at point $x$), and the
probabilities introduced above are functionals of this density, as well
as functions of the rapidity variable.

The average $S$--matrix is then computed as the following, functional,
integral:
 \beq\label{eq-Sfact}
    \avg{S}_Y =
    \int \mcal{D}n \mcal{D}m\, P_{\rmR}[n(x_{\rmR}),Y-Y_{\rm 0}]\,
    P_{\rmL}[m(x_{\rmL}),Y_{\rm 0}]\, S[n(x_{\rmR}),m(x_\rmL)],
   \eeq
which represents the average over all possible configurations of the
`event--by--event' $S$--matrix $S[n(x_{\rmR}),m(x_\rmL)]$ associated with
a given pair of configurations. In turn, the latter is given
by\footnote{Perhaps it is easier to understand this
    expression if we start from a discretized version where one has
    $S[n,m] = \prod_{ij} \sigma_{ij}^{n_i m_j}$ and then take the
    continuous limit to arrive at Eq.~\eqref{eq-Sgiven}.}
    \beq\label{eq-Sgiven}
     S[n,m] = \exp\left[\int \dif x_\rmR \dif x_\rmL n(x_\rmR)
    m(x_\rmL) \ln \sigma(x_\rmR|x_\rmL)\right], \eeq
where $\sigma(x_{\rmR}|x_{\rmL}) = 1 - \tau(x_{\rmR}|x_{\rmL})$ is the
$S$--matrix for the scattering of two elementary particles of logarithmic
sizes $x_{\rmR}$ and $x_{\rmL}$, and $\tau(x_{\rmR}|x_{\rmL})$ the
corresponding $T$--matrix (a real quantity, in between 0 and 1).

Consider now the evolution of any of the incoming systems with increasing
rapidity. An `evolution step' corresponds to a small increment $\dY$, and
we shall assume that only one extra particle can be emitted in such a
step. Moreover, the evolution law will be taken to be quite simple: when
one particle is emitted, the final configuration consists in the same
particles as the initial configuration plus an additional particle of
arbitrary size. We can quantify this assumption in terms of the following
master equation (with $\delta_{xz} \equiv \delta(x-z)$) :
 \beq\label{eq-master} \frac{\dif P[n(x),Y]}{\dY} = \int\limits_z
 f_z[n(x)-\delta_{xz}]\,P[n(x)-\delta_{xz},Y] -\int\limits_z
 f_z[n(x)]\,P[n(x),Y].\eeq
The quantity $f_z[n(x)]$ is a ``deposit'' rate density, that is
$f_z[n(x)] \dif z \dY$ is equal to the probability that we will find an
extra particle with logarithmic size in the interval $(z, z + \dif z)$
after one evolution step $\dY$, given that the initial configuration was
$n(x)$. The interpretation of the two terms in Eq.~\eqref{eq-master} as
gain (for the positive one) and loss (for the negative one) terms is then
straightforward.

Lorentz invariance (i.e., the condition that Eq.~\eqref{eq-Sfact} be
independent of $Y_0$) then requires \cite{ISST07} the deposit rate to be
proportional to $T_w[n(x)]$ --- the $T$--matrix for the scattering of a
particle of logarithmic size $w$ off a system at a given configuration
$n(x)$ :
   \beq\label{eq-Tgen}
     T_w[n(x)] = 1 - \exp\left[\int \dif x\, n(x) \ln \sigma(w|x)\right].
   \eeq
We fix the proportionality constant by choosing
  \beq\label{eq-fsimple}
    f_z[n(x)] = \frac{T_z[n(x)]}{\alpha(z)}.
   \eeq
where $\alpha(z)$ is the {\em running coupling} in the problem. This
precise dependence of the deposit rate upon $\alpha$ is chosen in such a
way to recover the correct powers of $\alpha$ in the evolution equations
to follow, in agreement with perturbative QCD. It is important to notice
that Eq.~\eqref{eq-fsimple} together with Eq.~\eqref{eq-Tgen} for the
scattering amplitude imply that the deposit rate density is in general a
highly non--linear function of the particle density. In turn, this means
that the extra particle at $z$ is emitted \emph{coherently} from all the
preexisting particles in the system.

To completely specify the model, we also need to specify the form of the
elementary particle--particle scattering amplitude $\tau(x|y)$ and the
$x$--dependence of the running coupling $\alpha(x)$. The former will be
chosen as
   \beq\label{eq-tau}
     \tau(x|y) = \alpha(x) \alpha(y) \exp(-|x-y|) \equiv
     \alpha(x) \alpha(y) K(x,y),
   \eeq
which mimics the corresponding quantity in QCD, i.e.~the elementary
dipole-dipole scattering amplitude at zero relative impact parameter.
Indeed, to lowest order in perturbative QCD (where the coupling is
fixed), the dipole--dipole amplitude (at zero impact parameter) is
approximately proportional to $\alpha^2 r_{<}^2/r_{>}^2$, where $r_< =
{\rm min}(r_1,r_2)$ and $r_> = {\rm max}(r_1,r_2)$; this is in agreement
with Eq.~\eqref{eq-tau} once one recalls that $x$ corresponds to the
logarithm of the inverse dipole size. Of course, after including running
coupling effects, which are part of the NLO corrections, the actual QCD
amplitude becomes more complicated than its toy model counterpart in
Eq.~\eqref{eq-tau}. Here however we shall stick to Eq.~\eqref{eq-tau},
which is general enough to illustrate our point, and has also the
advantage to be symmetric under the interchange $x \leftrightarrow y$.

As far as the running coupling $\alpha(x)$ is concerned, we would
naturally define it as $1/\beta x$, where $\beta$ is a fixed number
--- the analog of the one--loop beta--function in QCD. Such a
definition would introduce a divergence at $x\to 0$, similar to the
infrared problem of perturbative QCD. However, in the context of the
non--linear evolution of interest for us here, such a divergence is in
fact innocuous, except perhaps at low energies, because the {\em
saturation momentum} introduces an effective infrared cutoff, which grows
rapidly with $Y$ (see Sect. \ref{sect-analytic}). Still, in view of the
numerical simulations, it is necessary to regulate the singularity of the
running coupling at $x\to 0$; the onset of saturation will then guarantee
that the physical results are independent upon the precise regularization
prescription (at least, at sufficiently large $Y$). In what follows, we
choose to ``freeze'' the coupling when entering the ``non-perturbative''
region at $x\le 0$. (Note that our present conventions are such that
$x=0$ corresponds to a transverse momentum $k_\perp^2=\Lam^2$ in QCD.)
Namely, we shall require
   \beq\label{eq-alphagen}
    \alpha(x) =
    \begin{cases}
    {1/\beta x} \quad \!&\text{for}\quad x \gg 1 \\
    \order{1} \quad \!&\text{for} \quad x \sim \order{1}\\
    \alpha_0 <1 \quad \!&\text{for} \quad x = -\infty,
    \end{cases}
   \eeq
with $\alpha(x)$ a monotonically decreasing function of $x$. A precise
implementation of this running will be presented in Sect.
\ref{sect-numeric}.

Let us now return to Eq.~\eqref{eq-fsimple} for the deposit rate and
study its behavior in the limits of low and high particle density. We
easily find that
   \beq\label{eq-flimit}
    f_z[n(x)] \simeq
    \begin{cases}
    \int \dif x \,\alpha(x) K(z,x) n(x)
    &\text{when}\quad  n(x) \ll 1/\alpha(x)\alpha(z) \quad \text{for all }\, x\\
    1/\alpha(z) &\text{when} \quad n(x)
    \gg 1/\alpha(x)\alpha(z) \quad \text{for some }\, x.
    \end{cases}
   \eeq
Therefore, when the system is dilute the extra particle at $z$ is
(approximately) incoherently emitted from any of the preexisting
particles. This limit, where the deposit rate is proportional to the
particle density, and of $\order{\alpha_s}$, is analogous to the QCD
dipole picture \cite{AM94}. When the particle density starts to increase
the extra particle is emitted coherently, and in the limit where the
system becomes very dense the deposit rate saturates, at a value of
$\order{1/\alpha_s}$, and thus becomes independent of $n$. This is again
analogous to the situation encountered in QCD, where the JIMWLK equation
leads to the saturation of the gluon emission rate \cite{SAT}.

To conclude the presentation of the model, let us show how to construct
evolution equations for observables. In what follows, we shall consider
two types of observables: the particle density and the scattering
amplitude for a collision in which the left mover consists in a fixed
number of particles (meaning that the whole evolution up to rapidity $Y$
is given to the right moving system\footnote{Given the boost--invariance
of our formalism, this choice brings no loss in generality in so far as
the evolution is concerned, but merely amounts to specifying the initial
conditions for the left mover.}). These observables can be cast in the
form
  \beq\label{eq-Oave}
   \avg{\mcal{O}}_Y = \int \mcal{D}n\,
   P[n(x),Y]\, \mcal{O}[n(x)],
  \eeq
with the understanding that $P[n(x),Y]\equiv P_R[n(x),Y]$ for the
scattering problem. Differentiating this equation w.r.t. $Y$ and making
use of the master equation (\ref{eq-master}) one finds
  \beq\label{eq-Oevol}
   \frac{\del \avg{\mcal{O}}_Y}{\del Y} =
   \int \dif z\, \avg{f_z[n(x)]\,
   \left\{\mcal{O}[n(x)+\delta_{xz}] - \mcal{O}[n(x)]\right\}}.
  \eeq
In particular, the evolution of the average particle density is governed
by (dropping the index $Y$ and using subscripts instead of arguments from
now on)
   \beq\label{eq-nevol}
    \frac{\dif \lan n_x \ran}{\dY} =
    \avg{f_x[n]} = \frac{1}{\alpha(x)}\avg{T_x[n]},
   \eeq
with the natural interpretation that the rate of change in the average
particle density at $x$ is equal to the average deposit rate density at
$x$. In general, this is a complicated non-linear equation, which
involves (upon the expansion of the r.h.s.) all the $k$-body particle
densities, like the particle {\em pair} density ($k=2$), etc. But at low
density, Eq.~\eqref{eq-nevol} reduces to a linear equation
   \beq\label{eq-nBFKL}
    \frac{\dif \lan n_x \ran}{\dY} =
    \int\limits_z \alpha_z K_{xz} \avg{n_z}
    \quad \text{for} \quad n \ll 1/\alpha^2\,,
   \eeq
analogous to the running coupling version of the BFKL equation
\cite{BFKL} for the dipole density \cite{AM94}.

Consider now the $T$--matrix $T_x=1-S_x$ for the scattering of a single
left moving particle of logarithmic size $x$ off a generic right moving
system. For a given configuration of the latter, this is given by the
general expression Eq.~\eqref{eq-Sgiven} with $m(x_{\rmL}) =
\delta(x_\rmL - x)$. Making use of the general evolution equation
\eqref{eq-Oevol} we easily arrive at
    \beq\label{eq-Tone}
    \frac{\del \avg{T_x}}{\del Y} =
    \alpha_x \int\limits_z K_{xz} \avg{T_z(1-T_x)}
   \eeq
an equation analogous to the first Balitsky equation (extended to running
coupling). Once again, this is not a closed equation, as it involves the
$T$--matrix $\avg{T_x T_z}$ for the scattering of a projectile made with
two particles. It thus becomes necessary to write down the corresponding
evolution equation, which is similarly found as \beq\label{eq-Ttwo}
   \frac{\del \avg{T_x T_y}}{\del Y} \,&=&\,
   \alpha_x\int\limits_z K_{xz} \avg{T_z T_y (1-T_x)}+
   \alpha_y \int\limits_z K_{yz} \avg{T_z T_x (1-T_y)} \nn
   \,&+&\,\alpha_x \alpha_y \int\limits_z \alpha_z K_{xz} K_{yz}
   \avg{T_z (1-T_x)(1-T_y)},
  \eeq
This equation is analogous to the second Pomeron loop (PL) equation
\cite{IT04,IT05}. Notice that it is the last term, proportional to
$\alpha^3$, in the r.h.s. of the above equation which distinguishes this
equation from the corresponding one in the analogous B--JIMWLK hierarchy.
At a first glance, this last term seems to be suppressed with respect to
the first two terms when counting powers of $\alpha$. However this is not
true, since this term becomes a dominant one for $T\sim\alpha^2$. Let us
briefly discuss two limits of the hierarchy starting with the above
equations. While this hierarchy, as it stands, is not consistent with a
solution of the factorized form $\avg{TT...T} =
\avg{T}\avg{T}...\avg{T}$, it becomes so when we keep only the terms
which are explicitly proportional to $\alpha$. This is the mean field
approximation (MFA) in which the whole hierarchy reduces to the
factorized form of the first equation \eqref{eq-Tone}, that is,
  \beq\label{eq-TMFA}
    \frac{\del \avg{T_x}}{\del Y} =
    \alpha_x \int\limits_z K_{xz} \avg{T_z}(1-\avg{T_x}),
   \eeq
which is analogous to BK equation with running coupling. If we further
drop the nonlinear term proportional to $\avg{T}^2$ we arrive at the
toy--model analog of the BFKL equation for the dipole scattering
amplitude. Notice that in this case the running coupling stands outside
the integral in the r.h.s., to be contrasted with Eq.~\eqref{eq-nBFKL}
for the particle density where one needs to integrate over the
$z$--dependence of the coupling. As it should be clear from the previous
examples, in the present model it is the parent particle which sets the
scale for the argument of the coupling.

\section{From known results to expectations}\label{sect-analytic}
\setcounter{equation}{0}

In this section, we shall summarize the known results concerning the
fixed coupling case (with or without fluctuations) as well as the running
coupling case in the mean--field approximation. We shall specialize these
results to the one--dimensional model under consideration, but one should
keep in mind that similar results hold in QCD as well, under the
corresponding approximations. Based on such previous results, we shall
then formulate some theoretical expectations for the full problem
--- the stochastic evolution with running coupling.

The most important property of the non--linear evolution is the emergence
of the saturation `momentum' $x_s(Y)$, which is the scale separating
between the dense and dilute regions, and which increases with $Y$. This
scale is an intrinsic property of an evolving particle system, but it
also acts as the unitarization scale for the scattering between that
(evolving) system and an external dipole. For sufficiently large $Y$, the
scattering amplitude $T(x,Y)$ takes the form of a {\em front} which
interpolates between a strong scattering (saturated target) region at $x
\lesssim x_s$, where $T=1$, and a weak scattering (dilute target) region
at $x \gtrsim x_s$, where the amplitude decreases exponentially. (For the
stochastic evolution, this front picture holds event--by--event.) One can
then define the saturation line $x_s(Y)$ as the `position of the front',
that is, the line along which the amplitude is constant and of
$\order{1}$, e.g.~$T(x=x_s(Y),Y)=1/2$.

\vspace*{.2cm} \texttt{(i) }In the {\em fixed coupling} case, the
energy--dependence of $x_s$ is rather well understood, both in the MFA,
and in the full evolution including fluctuations.

\texttt{(i.a)} In the {\em mean--field} case, and for sufficiently large
$Y$, the `front velocity'
  \beq\label{eq-lambda}
   \lambda_s(Y) \equiv \frac{1}{\alpha}
   \,\frac{\dif x_s(Y)}{\dY} \,\approx\,
   \frac{\chi(\gamma_s)}{\gamma_s} - \frac{3}{2 \gamma_s \alpha Y} =
   3 \sqrt{3} - \frac{3 \sqrt{3}}{2 \alpha Y}\,,
  \eeq
slowly approaches the constant, asymptotic, value $\lambda_0\equiv
{\chi(\gamma_s)}/{\gamma_s} =3 \sqrt{3}$ from below.  Here $\chi(\gamma)
= 2/(1-\gamma^2)$ is the eigenvalue function of the linear equation
\eqref{eq-nBFKL} for fixed coupling and $\gamma_s = 1/\sqrt{3}$ solves
the equation $\chi(\gamma_s) - \gamma_s \chi'(\gamma_s) = 0$. Note that,
for fixed coupling (FC), it is the combination $\alpha Y$ which is the
natural evolution `time'. Furthermore, the amplitude in the {\em front
region} (the region ahead of the saturation line but relatively close to
it) is obtained as
   \beq\label{eq-satsol}
   T(x,Y) =
   c_1 (x-x_s+c_2) \exp\left[
   -\gamma_s (x-x_s) - \frac{(x-x_s)^2}{2 \chi''_s\alpha Y}
   \right],
  \eeq
valid for $1 \ll x-x_s \ll 2 \chi''_s\alpha Y$. In Eq.~(\ref{eq-satsol}),
$c_1$ and $c_2$ are unknown constants of $\order{1}$, and $\chi''_s\equiv
\chi''(\gamma_s) = 27$. We further notice that within the more restricted
window
   \beq\label{eq-winscalefc}
     1 \,\lesssim\, x-x_s\, \ll\,
      x_{\rm diff}(Y)\equiv \sqrt{2 \chi''_s \alpha Y}\,,
   \eeq
where the diffusion term in the exponent in Eq.~(\ref{eq-satsol}) can be
neglected, the amplitude exhibits {\em geometric scaling}, i.e. it
depends upon the kinematical variables $x$ and $Y$ only via the
difference $z\equiv x-x_s(Y)$.

For even larger $x$, such that $x-x_s\gg 2 \chi''_s\alpha Y$, the
amplitude shows `color transparency', i.e. it exhibits a faster,
exponential, decrease with $x$, which is the same as the long--range
decay of the elementary amplitude in Eq.~\eqref{eq-tau} : $T(x) \propto
\rme^{-x}$.

\texttt{(i.b)} After including {\em fluctuations}, one must distinguish
between the {\em event--by--event front}, corresponding to an individual
realization of the stochastic evolution, and the {\em statistical
ensemble of fronts}, which determines the average quantities.

\comment{
\begin{figure}[h]
    \centerline{
 \includegraphics[width=1.\textwidth]{FC_front_n_0.1.eps}}
  \vspace*{-.5cm}
    \caption{\small Fixed--coupling evolution in the one--dimensional
   model, for $\alpha=0.1$:
   the front corresponding to the {\em average} occupation number
   $\avg{n(x,Y)}$ is
   shown for different values of $Y$, starting from an initial condition
   $n(x,0)=n_0\Theta(x_0-x)$, with $n_0=2$ and $x_0=0$.}\label{fig:fcn}
\vspace*{.2cm}
\end{figure}}

\begin{figure}[t]
    \centerline{
    \includegraphics[width=7.4cm,angle=-90]{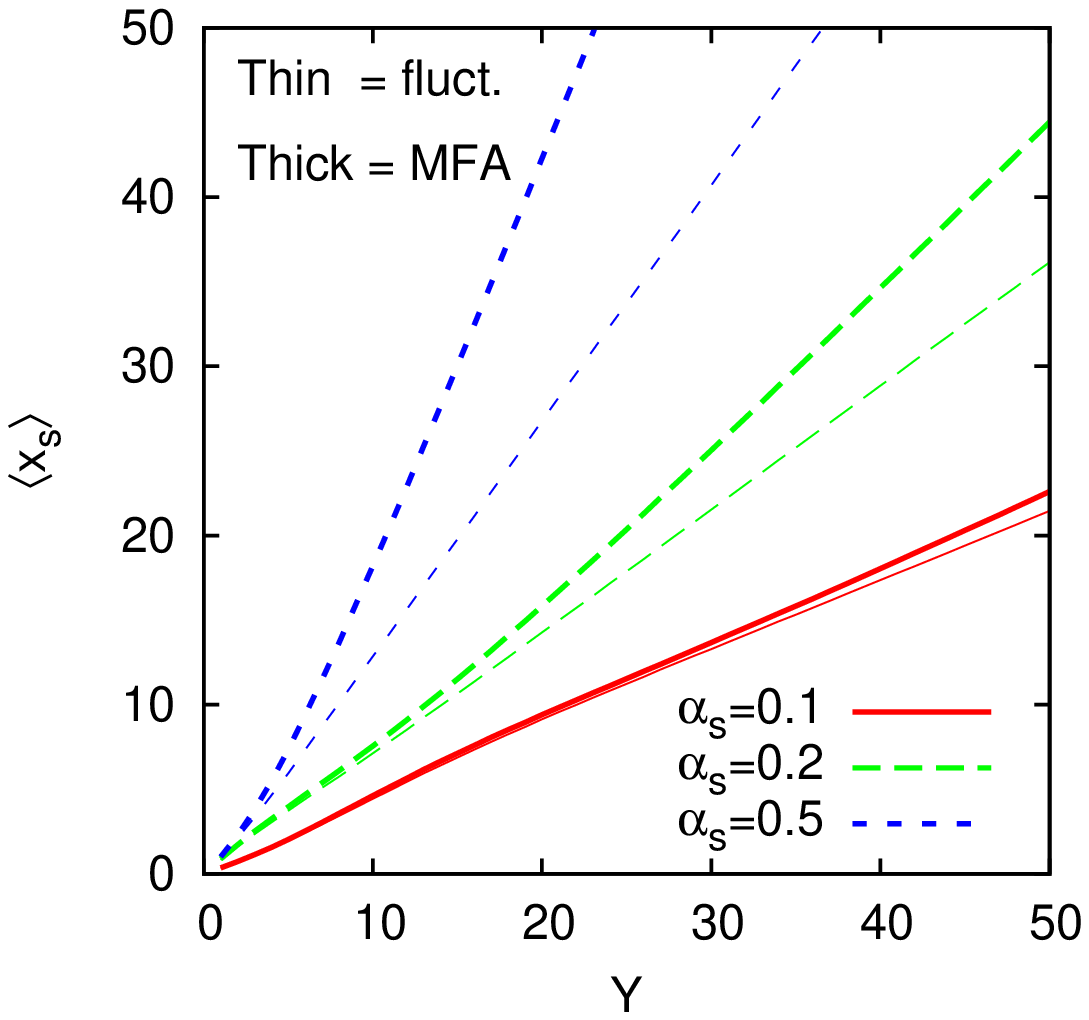}
    \includegraphics[width=7.4cm,angle=-90]{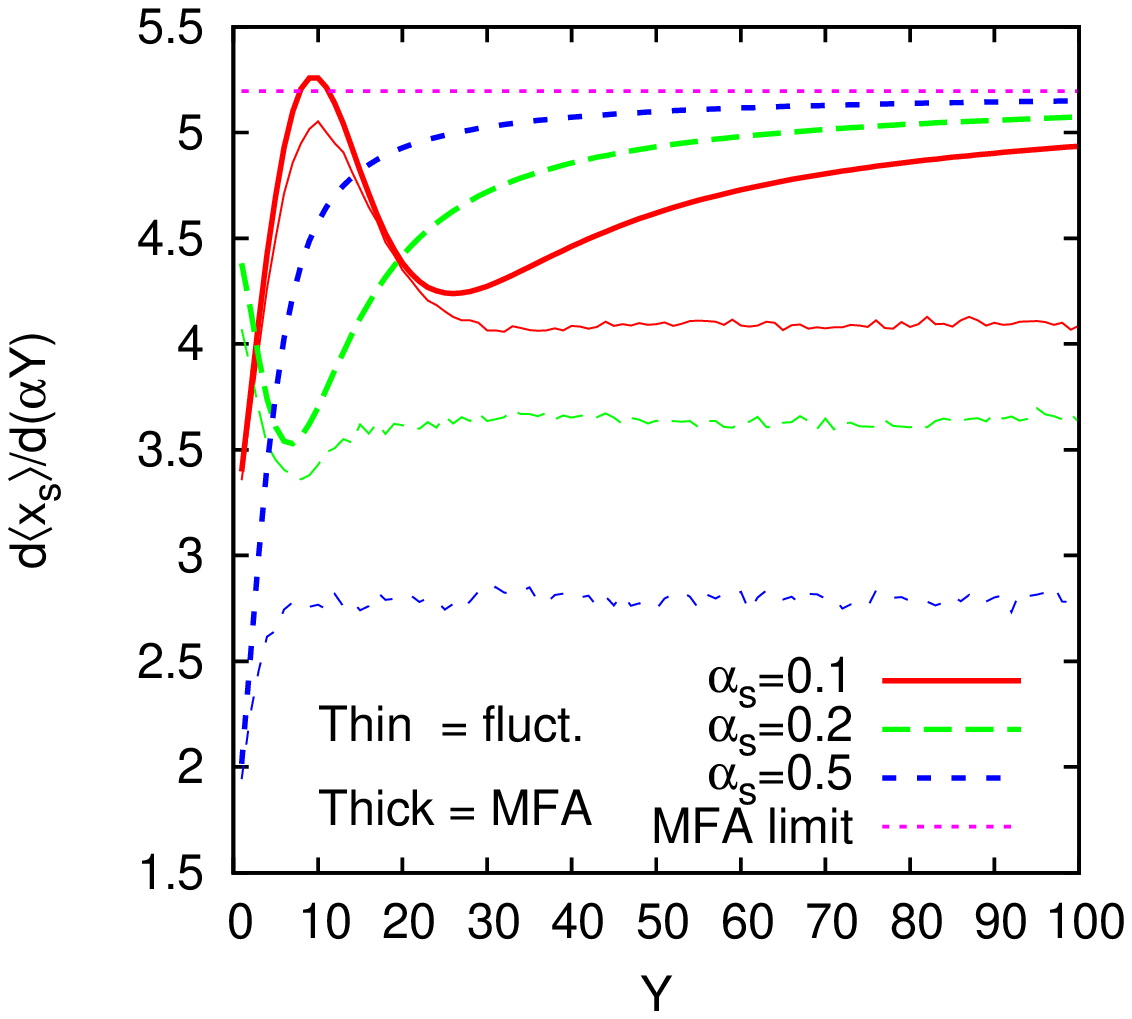}}
    \caption{\small The (average) saturation scale and the
    corresponding velocity in the
    evolution at fixed coupling, as obtained via the numerical
    study of the one--dimensional model, for 3 values of $\alpha$.
    The results of the full evolution (thin lines)
    are compared to the respective predictions of the MFA (thick lines).
     }\label{fig:fcxs}
\end{figure}

Up to fluctuations, the asymptotic velocity $\lambda_s$ is the same for
all fronts, and thus is also the same as the {\em average} (asymptotic)
velocity. But in contrast to the mean--field scenario, this asymptotic
value is now approached {\em much faster} (exponentially in $\alpha Y$),
and its value is {\em substantially smaller} than $\lambda_0$. Numerical
calculations show that the stronger $\alpha$ is, the larger is the
reduction in the value of $\lambda_s$ (see Fig.~\ref{fig:fcxs}; our
procedure for extracting the front position from the numerical results
will be explained in Sect.~\ref{sect-numeric}). An analytic estimate for
this reduction is known only in the limit where $\alpha$ is extremely
small, in which case one finds %
   \beq\label{eq-Afactor}
    \kappa \equiv \frac{\lambda_s}{\lambda_0}\,
    \simeq\,
    1 - \frac{\pi^2 \gamma_s^2 \chi''_s}{2 \chi(\gamma_s)
     \ln^2\alpha^2} =1 - \frac{3 \pi^2}{2 \ln^2 \alpha^2}\,.
   \eeq
This reduction in the value of the speed is in turn associated with a
profound change in the nature of the front, which is now {\em compact}.
This property is most transparent in the discretized version of the
model, as appropriate for numerical simulations: then, the number of
particles per bin is discrete, and thus it cannot be at the same time
non--zero and less than one. Hence, the (event--by--event) front
associated with the particle occupation number $n(x,Y)$ cannot have an
infinitely extending exponential tail, but rather it suddenly ends at
some `foremost occupied bin' $x_0$. The front for $T(x,Y)$, on the other
hand, is not truly compact, since an exponential tail
$T(x)\sim\alpha^2\rme^{-(x-x_0)}$ is generated at $x > x_0$ by the
scattering off the particles located around $x_0$. But in this case the
`compactness' refers to the fact that the front width $x_0-x_s$ cannot be
larger than $L \simeq (1/\gamma_s) \log(1/\alpha^2)$, which is the
distance over which the amplitude decreases from its value $T=1$ at
saturation down to a value $T \sim \mathcal{O}(\alpha^2)$. When this
maximal value is reached, one says that 'the front has been formed'.
Prior to that, the width of the front grows via diffusion, $x-x_s \propto
\sqrt{Y}$, so the front formation requires a typical 'formation time'
\beq\label{Dtau}
 \alpha \,Y_{\rm form}
 \simeq \,\frac{L^2}{2\chi''_s}\,
 \simeq\,
 \frac{\ln^2 \alpha^2}{2\chi''_s\,\gamma_s^2}\,.\eeq
To summarize, for fixed coupling and $Y > Y_{\rm form}$, the
event--by--event front for the scattering amplitude has the following,
approximate, shape, up to fluctuations  :
   \begin{equation}\label{TWD}
    T(x,Y)=
    \begin{cases}
        \displaystyle{1} &
        \text{ for\,  $z < 0$}
        \\*[0.2cm]
        \displaystyle{A\,z\,{\rm e}^{-\gamma_s z}
        \, 
        } &
        \text{ for\,  $1 < z < L$}
        \\*[0.2cm]
        \displaystyle{
        B\alpha^2{\rm e}^{- (z-L)}} &
        \text{ for\,  $z \gg L $},
    \end{cases}
 \end{equation}
with constant factors $A$ and $B$ of $\order{1}$. The precise
interpolations between the shown regimes, as well as the shape of the
front at earlier stages, $Y< Y_{\rm form}$, when the front has not yet
fully formed, are not under analytic control. Note also that, under the
present approximations, the event--by--event amplitude shows geometric
scaling: $T(x,Y)\approx T(x-x_s(Y))$.

\begin{figure}[t]
    \centerline{
    \includegraphics[width=7.4cm,angle=-90]{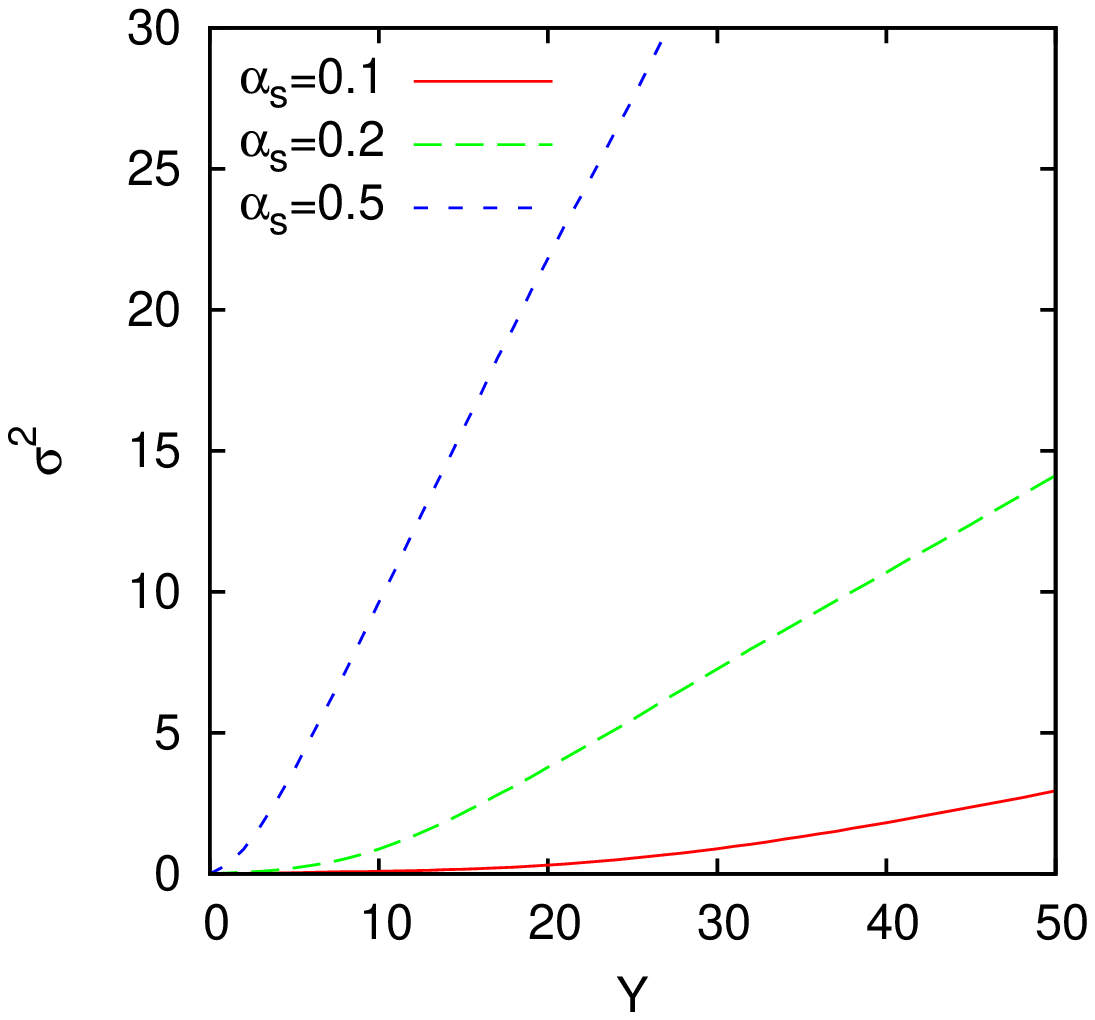}
    \includegraphics[width=7.4cm,angle=-90]{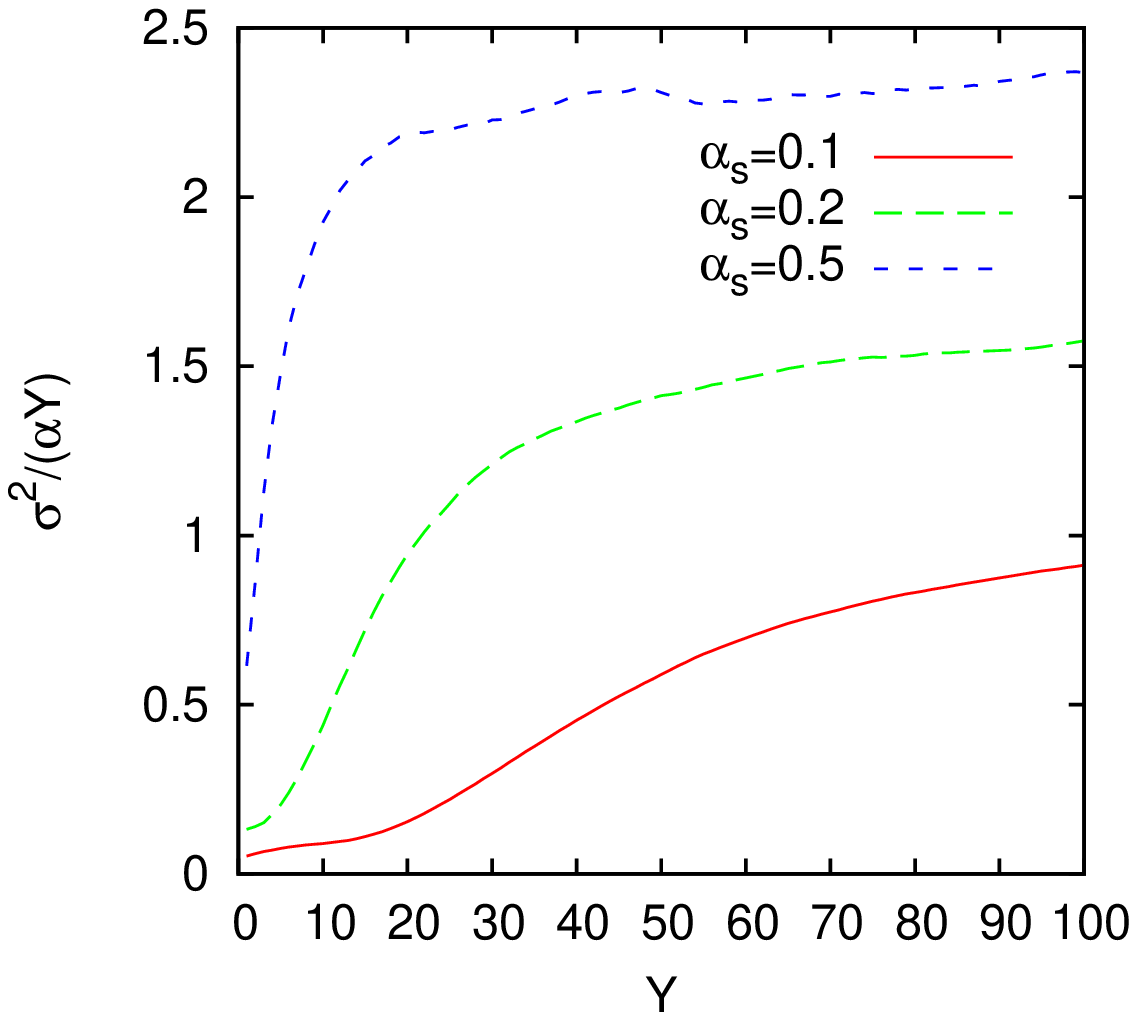}}
    \caption{\small
    The front dispersion in the fixed--coupling evolution
    for 3 values of $\alpha$; the respective values of the
    diffusion coefficient $D$ can be read off the figure on
    the right. Note the `formation time' $Y_{\rm form}$
    (which increases when decreasing $\alpha$)
    during which the dispersion remains negligible.
    }\label{fig:fcsigma}
\end{figure}

It turns out that the shape of the individual fronts is also important
for the {\em statistical} properties of the ensemble of fronts which is
obtained by repeating the same evolution a large number of times. The
fronts which compose this ensemble differ from each other via their
respective front position $x_s$, which now is a {\em random variable}.
The correspondence \cite{IMM04} with the reaction--diffusion process in
statistical physics suggests that, to a very good approximation, the
distribution of $x_s$ at $Y$ is a {\rm Gaussian}, with an expectation
value $\avg{x_s}$ and a dispersion $\sigma^2$ which both rise linearly
with $Y$ \cite{BDMM,MSX06}:
 \beq\label{eq-sigmaFC}
 \avg{x_s}(Y)=\lambda_s
\alpha Y\,,\qquad\sigma^2(Y)\equiv \avg{x_s^2}-\avg{x_s}^2 = D\alpha Y\,.
 \eeq
An analytic estimate for the {\em front diffusion coefficient} $D$ is
available only for $\alpha\to 0$, in which case one finds  \cite{BDMM} a
similar kind of logarithmic behavior as in Eq.~(\ref{eq-Afactor}):
 \beq\label{eq-DFC}
 D\, \simeq \,\frac{d_0}{\ln^3(1/\alpha^2)}\,,\eeq
with constant $d_0$. The very slow, {\em logarithmic}, convergence of the
results in Eqs.~(\ref{eq-Afactor}) and (\ref{eq-DFC}) to their respective
mean--field limits ($\kappa=1$ and $D=0$) as $\alpha\to 0$ reflects the
strong sensitivity of the fixed--coupling evolution to fluctuations. This
is confirmed by numerical simulations (in particular, for the model under
consideration \cite{ISST07}), which also show that, for more interesting
values of the coupling $\alpha$ (say $\alpha= 0.1\div 0.5$) the
coefficients $D$ and $1-\kappa=(\lambda_0-\lambda_s)/\lambda_0$ are
sizeable numbers, of $\order{1}$ (see Figs.~\ref{fig:fcxs} and
\ref{fig:fcsigma}). Hence, as manifest in Fig.~\ref{fig:fcsigma}, the
dispersion $\sigma^2(Y)$ rises quite fast with $Y$, which in turn has
profound consequences on the shape of the {\em average} amplitude
$\avg{T_x}$ \cite{IMM04,IT04} : when averaging over all the fronts in the
ensemble, the geometric scaling property characteristic of the individual
fronts, cf. Eq.~(\ref{TWD}), is washed out by dispersion and eventually
replaced, when $\sigma^2(Y)\simge 1$, by a new form of scaling known as
`diffusive scaling' \cite{HIMST06}\,: namely, $\avg{T_x}_Y$ scales as a
function of the single variable $(x-x_s(Y))/\sqrt{Y}$.

Moreover, the numerical simulations demonstrate a delay in the onset of
dispersion (in the plots in Fig.~\ref{fig:fcsigma}, this delay is
especially visible in the curves corresponding to smaller values of
$\alpha$), which can be roughly identified with the formation time
$Y_{\rm form}$. This relation between the formation time for the
individual fronts and the onset of dispersion is crucial in order to
appreciate our forthcoming results at running coupling, and can be
explained as follows: The growth in $\sigma^2$ is associated with the
rare fluctuations in which a new particle is produced relatively far away
ahead of the tip of the front (see, e.g., the discussion in Ref.
\cite{BDMM}). Such a fluctuation is most effective when the splitting
takes place near the tip of the front: first, this enhances the
probability that the daughter particle be produced further away
ahead\footnote{It has been argued in Ref. \cite{BDMM} that this
probability behaves like $P(\delta)\propto\rme^{-\gamma_s\delta}$ where
$\delta$ is the distance from the tip of the front $x_{\rm tip}\approx
x_s+L$ to the place where the new particle is generated.}; second, the
occupation numbers near the tip are low, of $\order{1}$, so the
fluctuations are relatively important there.

However, this whole picture starts to apply only {\em after} the front
has been formed, i.e., when the tail of the front $n(x,Y)$ has developed
the characteristic `anomalous dimension' $\gamma_s$ all the way down to
its tip. Assume that, at $Y_0=0$, one starts with the initial condition
$n(x,0)=n_0\Theta(x_0-x)$, with $n_0=\order{1}$. Then, in the originally
occupied bins at $x<x_0$, the occupation numbers will rise very fast with
$Y$, as $n(x,Y)\simeq n_0 \rme^{2\alpha Y}$ \cite{ISST07}, until they
reach a large, `saturation', value, of $\order{1/\alpha}$; this happens
around $ Y_{\rm sat}\simeq (1/2\alpha)\ln(1/\alpha)$, which at weak
coupling is parametrically smaller than $Y_{\rm form}$. Thus, during the
intermediate evolution at $Y_{\rm sat} \simle Y \simle Y_{\rm form}$, the
front $n(x,Y$) shows a very abrupt fall off (much faster than the
exponential law $\rme^{-\gamma_s(x-x_s)}$ to be eventually reached at
$Y\sim Y_{\rm form}$) from a saturation value $n_{\rm sat}\sim 1/\alpha$
down to $n=1$. Such an abrupt fall off is not favorable for developing
fluctuations, since the front has no distant tip with occupation numbers
of $\order{1}$); hence, during this interval in $Y$ the front evolution
remains `mean-field'--like. Fluctuations start to significantly develop
only for $Y> Y_{\rm form}$.

\vspace*{.2cm}
 \texttt{(ii)} We now turn to the {\em running coupling}
case, which so far has been investigated only  at the {\em mean--field}
level, i.e., at the level of the BK equation in QCD, and at different
levels of sophistication for implementing the running coupling effects
\cite{Motyka,SCALING,MT02,DT02,MP032,RW03,IIT04,Nestor04,GKRW,KWrun1,Brun,KWrun2,Alb07}.
For the present model, the corresponding, non--linear, equation in shown
in Eq.~\eqref{eq-TMFA}.

Such previous studies have shown that, for the purpose of computing the
high--energy asymptotic of the saturation scale, one can assume that the
front speed is determined \emph{locally} by the corresponding fixed
coupling result \cite{GLR,SCALING}; that is,
 \beq\label{eq-dxsat}
    \frac{\dif x_s}{\dY}\,
     \simeq \,\lambda_0\alpha(x_s) \simeq
    \frac{\lambda_0}{\beta x_s}\,
    \Longrightarrow\,
     x_s(Y) \simeq \sqrt{\frac{2 \lambda_0}{\beta}\,Y},
   \eeq
where we have used the asymptotic FC result in Eq.~\eqref{eq-lambda}
together with $\alpha(x_s)=1/(\beta x_s)$. This simple argument is
justified since the relevant dynamics takes place within a limited
distance ahead of $x_s$, of the order of the `diffusive radius' $x_{\rm
diff}$. The latter rises with $Y$, but this rise is considerably slower
than that of the saturation scale itself; thus, the {\em relative} width
$x_{\rm diff}/x_s$ of the active region decreases with $Y$. In fact, the
rise of the diffusive radius with increasing energy is much slower with
RC than with FC \cite{MT02,DT02,IIT04} (see below), and this difference
turns out to have crucial consequences for the physics of fluctuations,
as we shall discover later on.

Based upon the above considerations, it is possible to perform a more
accurate study \cite{MT02,DT02,MP032} of the solution to the mean--field
equation \eqref{eq-TMFA}, in which the argument $x$ of the running
coupling is self--consistently expanded around $x_s(Y)$. One thus finds
(up to a multiplicative constant)
    \beq\label{eq-tsolmft}
    T(x,Y) \,\simeq\,
    \exp[-\gamma_s (x-x_s)]\,
    \tau^{1/3}
    {\rm Ai}\left(\xi_1 + \frac{x-x_s + c}{D_s \tau^{1/3}}
    \right)
    \exp\left[
    -\frac{2 (x-x_s)^2 }{3 \chi''(\gamma_s) \tau}
    \right],
    \eeq
where the saturation line $x_s(Y)$ is more accurately determined by
    \beq\label{eq-lambdarc}
    \lambda_s \equiv \frac{\dif x_s}{\dY}\,\simeq\,
    \frac{\lambda_0}{\beta}
    \left(
    \frac{1}{\tau}
    -\frac{|\xi_1|D_s}{4}\,
    \frac{1}{\tau^{5/3}}
    \right).
    \eeq
Here we have defined $\tau \equiv \sqrt{(2 \lambda_0/\beta)(Y+Y_0)}$,
with $Y_0$ an unknown constant. Furthermore, $D_s\equiv
\{\chi''(\gamma_s)/[2 \chi(\gamma_s)]\}^{1/3}\approx 1.65$, Ai is the
Airy function, and $\xi_1 \simeq -2.34$ is the location of its rightmost
zero. As compared to the earlier estimate in Eq.~\eqref{eq-dxsat}, the
expression \eqref{eq-lambdarc} also provides the first pre--asymptotic
correction to the front velocity, to be compared to the corresponding FC
result in Eq.~\eqref{eq-lambda}.

The above formul\ae\ show that it is natural to interpret the variable
$\tau\propto \sqrt{Y/\beta}$ as the `evolution time' with RC.  For the
purposes of the numerical calculations in the next section, it is also
useful to define a `velocity' with respect to this natural `evolution
time'; we shall write
  \beq\label{eq-vrc}
    v_s \equiv \frac{\dif x_s}{\dif \sqrt{(Y/\beta)}}\,\simeq\,
    \sqrt{2\lambda_0}
    \left(1
    -\frac{|\xi_1|D_s}{4}\,
    \frac{1}{\tau^{2/3}}
    \right).
    \eeq
At large $Y$, this velocity approaches the constant value
$v_0=\sqrt{2\lambda_0}$ which is independent of $\beta$.

The expression \eqref{eq-tsolmft} for the amplitude is valid in the
interval
   \beq\label{eq-eqwin}
     1\,\lesssim\, x-x_s \,\ll \,
     \frac{3 \gamma_s D_s^2}{\sqrt{5}}\,\tau^{2/3}\,,
   \eeq
which defines the `front region' for RC. When one restricts oneself to
the narrower window
   \beq\label{eq-winscale}
     1 \,\lesssim\, x-x_s \,\ll \, x_{\rm diff}(Y)\equiv
     D_s \tau^{1/3}\,,
   \eeq
then one finds that the amplitude shows a `geometric scaling' behavior,
   \beq\label{eq-tscale}
     T(x,Y) \,\simeq\,(x - x_s + c) \exp[-\gamma_s (x-x_s)],
   \eeq
formally similar to the FC case. Once again, the upper bound of the
geometric scaling region defines the diffusive radius. As anticipated,
with RC, this radius grows very slowly with $Y$\,: $x_{\rm diff} \sim
\tau^{1/3}\sim Y^{1/6}$, to be compared to the much faster increase at
FC\,: $x_{\rm diff} \sim Y^{1/2}$, cf. Eq.~\eqref{eq-winscalefc}. Some
numerical results for the mean--field evolution with running coupling
(RC) (which will confirm the above analytic estimates) will be presented
in Sect.~\ref{sect-numeric}.

We note that in order to obtain the above scaling in $Y$ of the diffusion
radius, it was rather crucial that we did {\emph not} set the scale
determining the argument of the coupling to be the saturation scale.
(Such an approximation would not be correct and would lead to a diffusion
radius proportional to $\sqrt{\tau}\sim Y^{1/4}$ which is smaller than
the actual one.) In Eq.~\eqref{eq-tsolmft}, we have two sources of
diffusion, namely, the Airy function, for which $x_{\rm diff} \sim
\tau^{1/3}\sim Y^{1/6}$, and the Gaussian function, for which $x_{\rm
diff} \sim \tau^{1/2}\sim Y^{1/4}$. In the overall product, it is the
smaller of the two diffusive radii that controls the physics; indeed, the
diffusion radius plays the role of the effective phase space for
evolution, and therefore the smaller it becomes, the larger the
preasymptotic corrections are\footnote{Notice that the relative
correction in \eqref{eq-lambdarc} is of order $\tau^{-2/3} \sim x_{\rm
diff}^{-2}$. This ``inverse--square law'' also holds in the FC case, as
manifest on Eq.~\eqref{eq-lambda}.}. Indeed, had we set the coupling
equal to $1/\beta x_s$, we would have found a smaller correction, of
order $1/\tau^2$, to \eqref{eq-lambdarc}.

\vspace*{.2cm}
 \texttt{(iii)}
Based on the previous discussion, we are now prepared to formulate some
{\em theoretical expectations} for the results of the full problem, which
includes {\em both} running coupling and particle--number fluctuations.
In the next section, these expectations will be confronted to the actual
numerical results.

Let us start with the {\em asymptotic} regime, where the situation looks
conceptually simpler. (We shall later specify when we expect this
asymptotic regime to install.) Then, by the same argument as discussed in
relation with Eq.~\eqref{eq-dxsat}, we expect the rate of change in both
the average saturation scale and the dispersion be governed by the
respective results at fixed coupling, cf. Eq.~(\ref{eq-sigmaFC})
evaluated with the {\em local} value of the coupling $\alpha(\avg{x_s})$.
That is:
 \beq\label{eq-dYfl}
   \frac{\dif \avg{x_s}}{\dY} \simeq\kappa\lambda_0\, \alpha(\avg{x_s})
   = \frac{\kappa\lambda_0}{\beta \avg{x_s}}\,,\qquad\ \
    \frac{\dif \sigma^2}{\dY} \simeq D\, \alpha(\avg{x_s}).
   \eeq
From the FC case, one should recall that the coefficients $\kappa$ and
$D$ in the above equations depend upon $\alpha$ (the stronger the larger
is $\alpha$), and hence they become $Y$--dependent in this RC case.
However, for sufficiently large values of $Y$, the coupling
$\alpha(\avg{x_s})$ becomes arbitrarily small, and then the estimates in
Eqs.~(\ref{eq-Afactor}) and (\ref{eq-DFC}) become reliable. These
formul\ae\ show that, in this asymptotic regime, the dependence of the
coefficients $\kappa$ and $D$ upon $\alpha$ (and hence upon $Y$) is
merely logarithmic and thus can be neglected in integrating
Eq.~\eqref{eq-dYfl}. One then finds:
 \beq\label{eq-sigmarun}
    \avg{x_s}\,\simeq\,\sqrt{\frac{2\kappa\lambda_0}{\beta}(Y+Y_0)}\,,
    \qquad\ \
     \sigma^2 \,\simeq\,
     D \,\sqrt{\frac{2}{\beta\kappa\lambda_0} (Y+Y_0)}\,,
   \eeq
where, strictly speaking, the factors $\kappa$ and $D$ have a weak
dependence upon $Y$, and the integration constant $Y_0$ is not under
control. Moreover, $\kappa$ should be strictly smaller than one, and it
should slowly approach one from below when $Y\to\infty$. In the same
limit, $D$ should slowly approach zero, and the front velocity $v_s$
defined as in Eq.~\eqref{eq-vrc} should approach the value
$v_0=\sqrt{2\lambda_0}$; this is the same limiting value as in the
mean--field case, but the convergence towards it should be much slower
with fluctuations (because of the slow convergence of $\kappa$ towards 1)
than in the MFA.

While the above considerations look reasonable indeed, they do not tell
us {\em how fast} is this universal high--energy regime approached when
increasing $Y$. This question is even more crucial at RC than it was at
FC, since with RC, the high--energy limit $Y\to\infty$ is tantamount to
the weak coupling limit $\alpha\to 0$. Hence, the most interesting
phenomena may be concentrated in the early (or `pre-asymptotic') stages
of the evolution, where the coupling is stronger.

In order to answer this question (at a qualitative level, at least), we
need to study the front `formation time' with RC. This is determined by
the same general argument as for FC, namely this is the rapidity
evolution $Y_{\rm form}$ which is required for the diffusive radius to
become as large as the width $L$ of an individual front. From
Eq.~\eqref{eq-tscale} (which remains approximately true for the
event--by--event fronts even in the presence of fluctuations), $L$ is
evaluated as:
 \beq\label{eq-LRC}
  L\,\simeq \,\frac{1}{\gamma_s}\ln\frac{1}{\alpha^2(x_s)}
  \,\simeq\,
  \frac{1}{\gamma_s} \ln \frac{Y}{\beta}\,,\eeq
and hence it is slowly increasing with $Y$. By also using the RC--version
of the diffusive radius $x_{\rm diff}$, as given by the upper limit in
Eq.~\eqref{eq-winscale}, one finds
 \beq
 x_{\rm diff}(Y_{\rm form})\,\simeq\,L\ \Longrightarrow\
 \ \frac{Y_{\rm form}}{\beta}\,\simeq\,\frac{1}{2\lambda_0}\left(
 \frac{L}{D_s}\right)^6\,.
  \eeq
This is, strictly speaking, a transcendental equation (since $L$ itself
depends upon $Y_{\rm form}$, albeit only slowly, cf. Eq.~\eqref{eq-LRC}),
that we shall not attempt to explicitly solve here, since our argument is
at best qualitative. Rather, for our purposes, it suffices to notice
that, with RC, the formation time scales like $Y_{\rm form}\sim \beta
L^6$, and hence it is parametrically larger than in the FC case (where we
have seen that $Y_{\rm form}\sim (1/\alpha)L^2$, cf. Eq.~\eqref{Dtau}).
In a previous discussion in this section, we have argued that the
formation time for the individual fronts also acts as the onset time for
the growth of dispersion. We thus conclude that, with RC, the onset of
the fluctuation effects, and thus of the universal, asymptotic, behavior
in the high--energy evolution, should be strongly delayed. Our previous,
analytic, estimates are too crude to more quantitatively characterize
this delay. We therefore turn to a numerical analysis, with results to be
presented in the next section.

 \section{Numerical results}\label{sect-numeric}
\setcounter{equation}{0}

In this section, we shall numerically study the evolution of the particle
distribution in a `hadronic' system described by our model (with running
coupling and fluctuations), and also the scattering between this system
(the `target') and a simple, unevolving, projectile, which consists in a
single particle of variable `size' $x$. We shall consider initial
conditions where the particle density shows a plateau at $Y_0=0$ :
$n(x,0)=n_0\Theta(x_0-x)$, with $n_0$ chosen in such a way that the
original {\em occupation numbers} (the number of particles per bin) be of
$\order{1}$. In practice, we divide the $x$ axis into bins of width
$\Delta x=1/8$ and start the evolution with 2 particles per occupied bin,
meaning $n_0=16$. (We have checked that our results are insensitive to
the discretization prescription.) This choice introduces a dissymmetry
between the target and the projectile already in the initial conditions:
the target is relatively denser, although still far away from saturation;
this dissymmetry will be, of course, further amplified by the subsequent
evolution.

In our simulations, we shall pay special attention to ensure that the
final results are insensitive to the prescription used to `freeze' the
coupling at negative values of $x$ (cf. Eq.~\eqref{eq-alphagen}). This is
necessary in order to unambiguously distinguish between the results of a
perturbative evolution with running coupling (which is the problem of
interest for us here) and the `non--perturbative' phenomena associated
with the freezing of $\alpha$. Clearly, this is an issue only for the
evolution in the {\em early stages} : for sufficiently large values of
$Y$, the saturation scale is large too, $x_s(Y)\gg 1$, and acts as a
`hard infrared cut-off' which removes any sensitivity to the
`non--perturbative' region\footnote{This should be contrasted to the
linear, BFKL--like, evolution, where the sensitivity to the
non--perturbative region at low momenta persists up to arbitrarily large
$Y$, because of the `infrared diffusion'.} at $x\le 0$. However, if the
freezing is important in the early stages, then this early evolution is
`fixed-coupling'--like, and thus it is relatively fast. Hence, the
results of the evolution accumulated in these early stages can be
numerically important and difficult to disentangle from those of the
running coupling evolution at later stages. To avoid this, we shall make
sure that, in the whole range of $x$ covered by the evolution, the
running coupling takes its perturbation shape $\alpha(x) = {1/\beta x}$
to a very good accuracy.

In practice, we shall use the following prescription for the running
coupling, which provides a smooth interpolation (actually, a family of
such interpolations) to the behavior in Eq.~\eqref{eq-alphagen}:
   \beq\label{eq-alpha}
    \alpha(x) =
    \frac{1}{\beta c\ln \left(\rme^{x/c} + \rme^{1/\alpha_0 \beta c}
    \right)}\,,
   \eeq
where $0 < c \le 1$. For $x> 1/(\beta\alpha_0)$, this running coupling
becomes independent of its value in the infrared, $\alpha_0$. The
additional parameter $c$ allows us to control how fast is this
transition, from a frozen coupling to a perturbative one: the smaller
$c$, the sharper the transition. We have checked that, with $c=0.1$,
there is no influence of freezing as soon as $x\simge 2/(\beta\alpha_0)$.

In the numerical calculations to be presented below, we have chosen
$\alpha_0=0.7$, $c=0.1$, and a value for $x_0$ (the upper bound of the
initial plateau at $Y=0$) which is related to $\beta$ in such a way that
the condition $x_0
> 2/(\beta\alpha_0)$ be satisfied. We have considered several values for
$\beta$ (the parameter which controls the speed of the perturbative
running), but payed special attention to the case $\beta=0.72$, which is
similar to the one--loop beta function of QCD (with $N_f=3$ flavors of
quarks). We have performed systematic calculations up to a maximal
rapidity $Y_{\rm max}=200$.

The fundamental quantity in our analysis is the position of the front (or
`saturation scale') $x_s(Y)$, which can be defined as a line of constant
amplitude for the dipole scattering --- say, $T(x=x_s(Y),Y)=0.01$. It
turns out, however, that a more convenient definition in practice is the
following one\footnote{We have checked that our numerical results are
unchanged if one instead defines $x_s$ using $T(x=x_s)=\text{cst.}$.}
($T(x,Y)$ is the event--by--event amplitude) :
   \beq\label{eq-xsat}
    x_s(Y) = x_s(0) + \int\limits_x \left[ T(x,Y) - T(x,0)\right],
   \eeq
which exploits the fact that for a given event the amplitude satisfies
geometrical scaling, i.e.~it is a function only of the combined variable
$x-x_s(Y)$, cf. Eq.~\eqref{TWD}. (This property is not exact, but it is
satisfied for all values of $x$ that give the dominant contribution to
Eq.~\eqref{eq-xsat}.) This definition together with Eq.~\eqref{eq-xsat}
implies the following representation for the front dispersion
   \beq\label{eq-disp}
     \sigma^2(Y) \equiv \lan x_s^2\ran - \avg{x_s}^2 =
     \int\limits_{xy} \left[\avg{T_x T_y} - \avg{T_x}\avg{T_y}
     \right],
   \eeq
where the expectation values in the r.h.s. are computed at rapidity $Y$.

In practice, of course, we shall extract $x_s(Y)$ for each individual
front via Eq.~\eqref{eq-xsat}, and then construct statistical quantities,
so like $\avg{x_s}$, $\sigma^2$, and also the 3rd order cumulant, $ C_3
\equiv \lan x_s^3 \ran -
    3 \lan x_s^2 \ran \lan x_s \ran + 2 \lan x_s \ran ^3$,
via a statistical analysis over the ensemble of events. (Notice that
$C_3$ would exactly vanish if the event distribution was Gaussian.)

\begin{figure}[t]
    \centerline{
    \includegraphics[width=7.4cm,angle=-90]{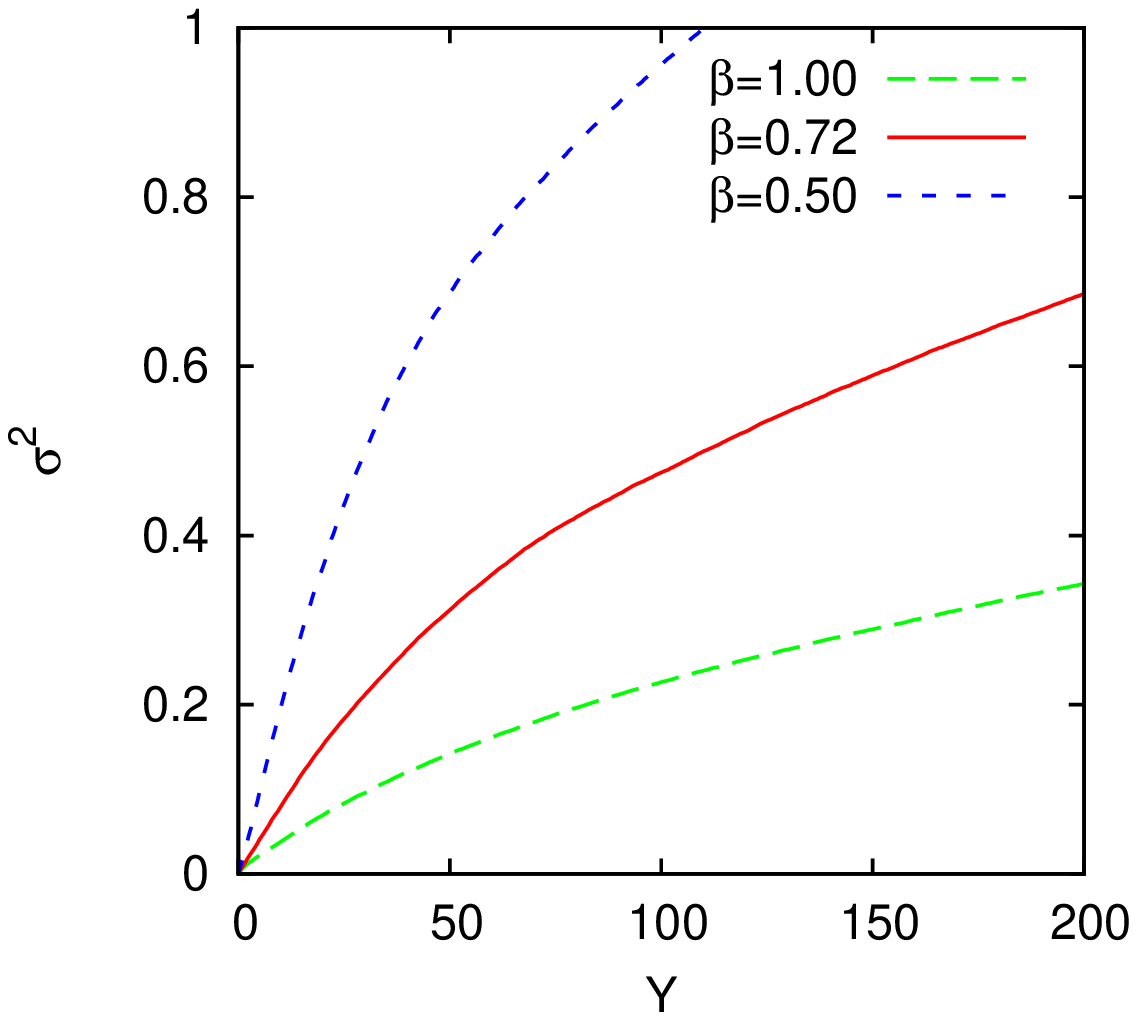}
    \includegraphics[width=7.4cm,angle=-90]{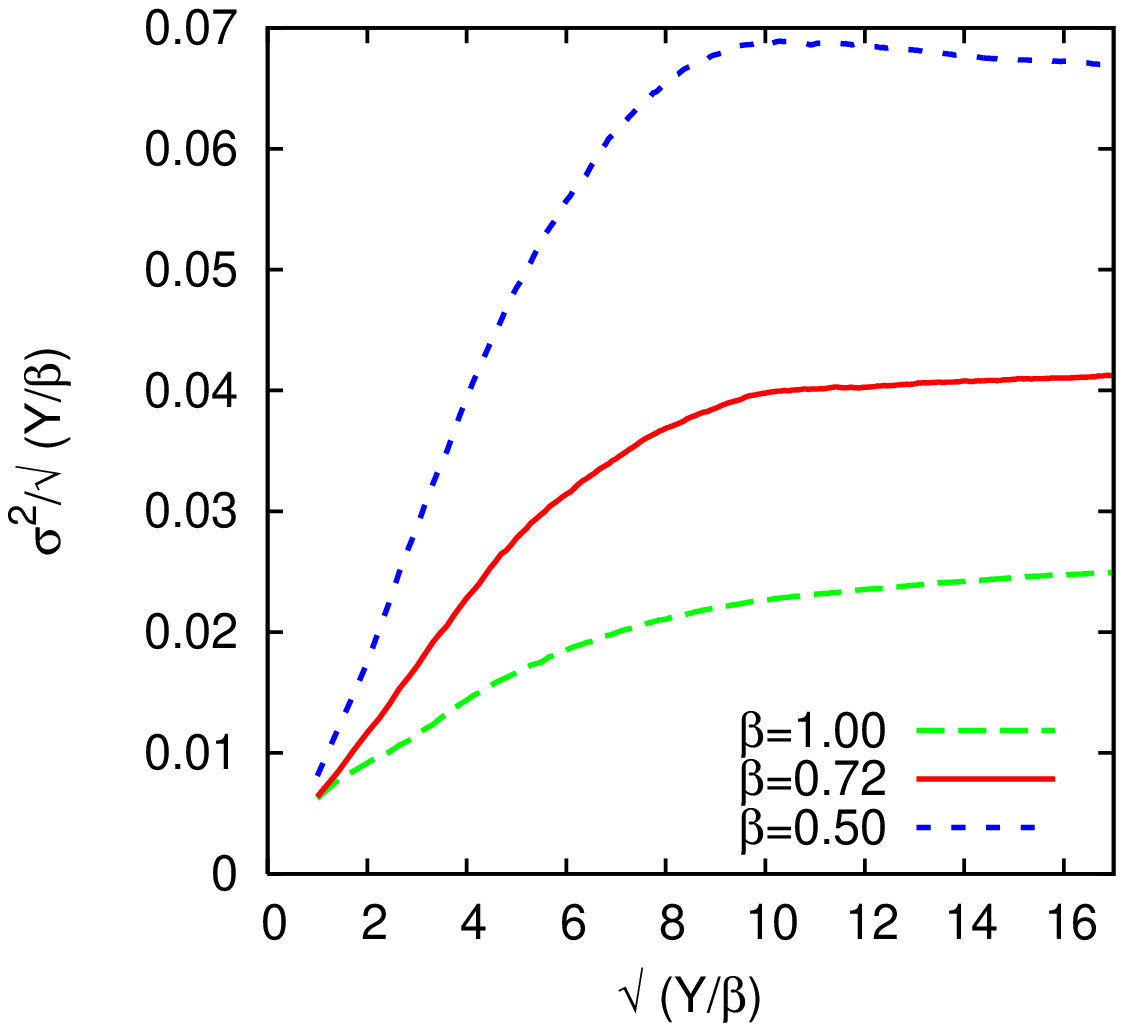}}
    \caption{\small
    The front dispersion in the evolution with running coupling,
    for $Y\le 200$ and 3 values of $\beta$.
    }\label{fig:rcsigma}
\end{figure}

We are now in a position to present our numerical results. The main
conclusion emerging from these results can be succinctly formulated as
follows: {\em the `Pomeron--loop' effects (i.e., the influence of the
particle--number fluctuations on the evolution towards saturation) are
strongly suppressed in the evolution with running coupling, and remain
negligible for all the rapidities of interest.} The physical origin of
this suppression is {\em the slowing down of the evolution by the running
of the coupling} and, more precisely, the large `formation time' required
for the formation of the front and the onset of fluctuations. This
conclusion, and its above physical interpretation, are supported by the
ensemble of the numerical evidence we now present.

\vspace*{.3cm}

\texttt{(i)} {\em The strong suppression in the dispersion $\sigma^2$ and
the 3rd cumulant $C_3$} \vspace*{.2cm}

In Fig. \ref{fig:rcsigma} we have displayed our numerical results for the
dispersion $\sigma^2(Y)$, for rapidities  $Y\le 200$ and two values of
the parameter $\beta$. The dispersion grows with $Y$, as expected;
moreover, as visible in the right hand plot, this rise is roughly
consistent (at least for sufficiently large values of $Y\simge 100$) with
the $\sqrt{Y}\,$--law expected according to Eq.~\eqref{eq-sigmarun}.

\begin{figure}[t]
    \centerline{
    \includegraphics[width=7.4cm,angle=-90]{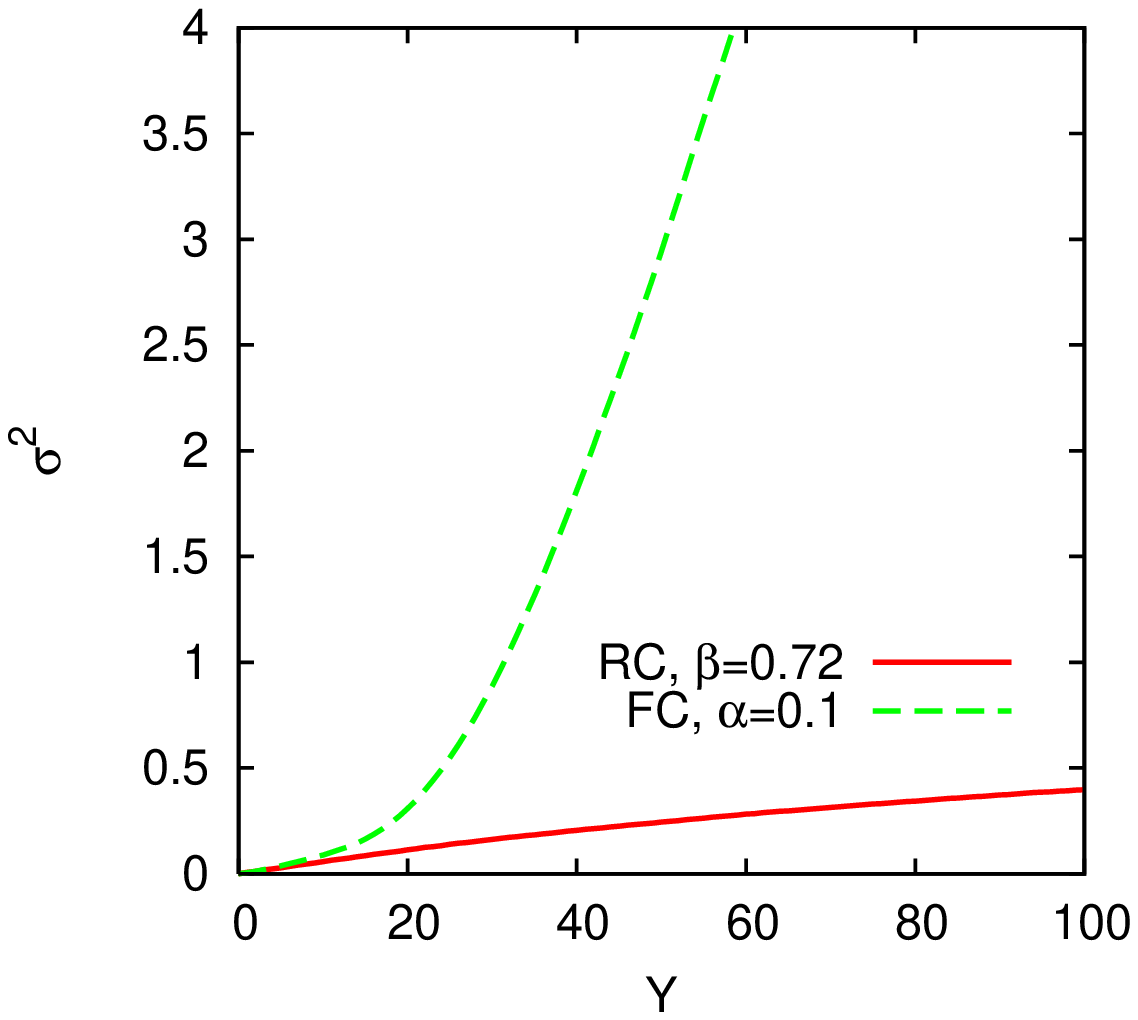}
    \includegraphics[width=7.4cm,angle=-90]{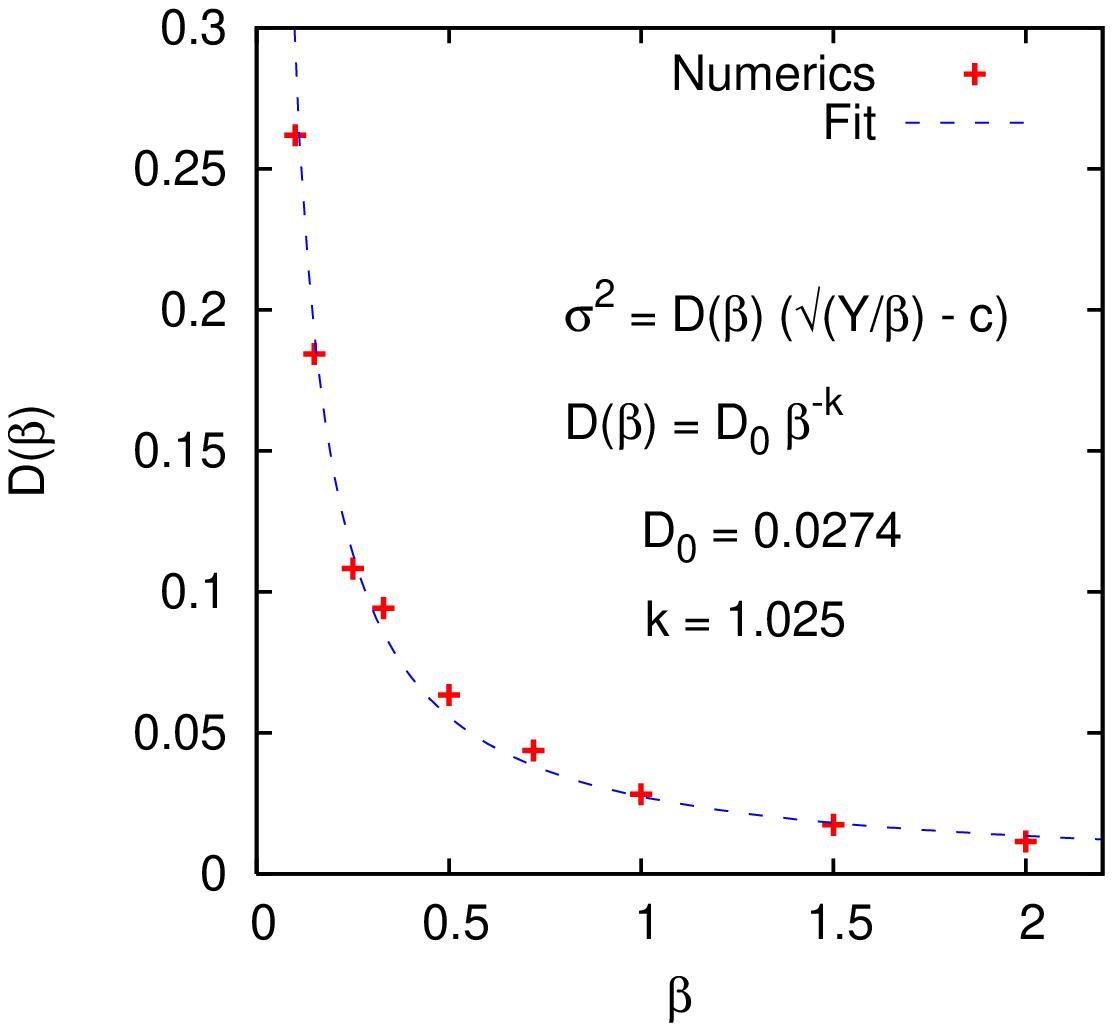}}
        \caption{\small Left: A comparison between FC and RC results
    for the dispersion. Right: The front diffusion coefficient with
    running coupling, as extracted from a fit to the numerical results.}
    \label{fig:frcsigma} 
\end{figure}

What is perhaps less expected, and thus surprising at a first fight, is
the {\em extremely small magnitude} of the measured dispersion, as
compared to the respective fixed coupling results (compare in this
respect with Fig. \ref{fig:fcsigma}). One might think that this reduction
in $\sigma^2$ is due to the fact that, with RC, the coupling is
effectively weaker, but this is actually {\em not} true. To demonstrate
this, we have compared in the left hand side of Fig. \ref{fig:frcsigma}
the dispersion $\sigma^2$ produced in the RC run with $\beta=0.72$ and,
respectively, the FC run with $\alpha=0.1$. This particular value
$\alpha=0.1$ is appropriate for our argument since it is close to the
{\em smallest} value of the running coupling reached in the corresponding
RC simulation; namely, this is the same as
$\alpha(\avg{x_s})=1/\beta\avg{x_s}$ with $\avg{x_s}$ measured at $Y=75$
(see below). Yet, as manifest on Fig. \ref{fig:frcsigma}, the total
dispersion accumulated in the RC evolution is tremendously smaller (by
almost two orders of magnitude) than the corresponding result with FC.

\begin{figure}[t]
    \centerline{
    \includegraphics[width=12.cm]{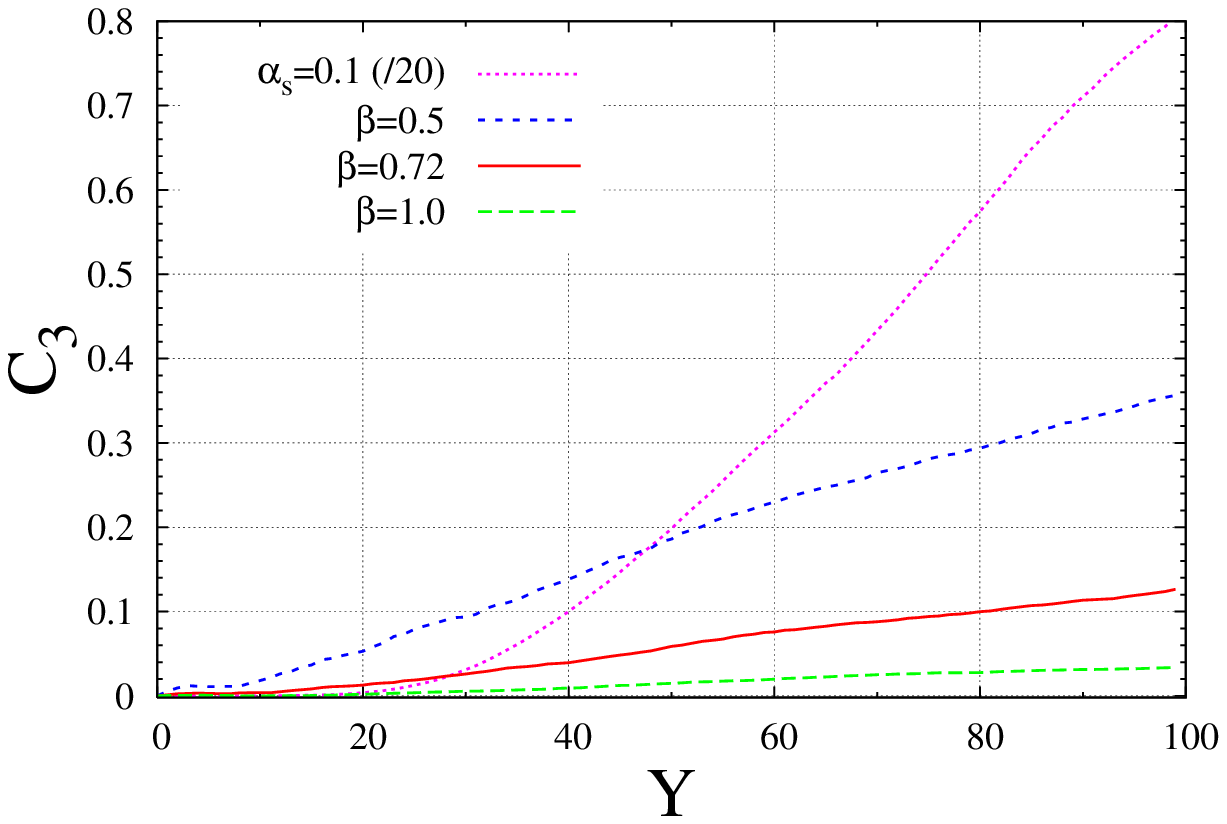}}
    \caption{\small
    The 3rd cumulant $C_3$ in the evolution with running coupling,
    for $Y\le 100$ and 3 values of $\beta$. For comparison, the
    fixed coupling result corresponding to $\alpha=0.1$ is also shown;
    note that, in order to fit inside this plot,
    the FC result has been divided by a factor of 20.
    }\label{fig:frcC3}
\vspace*{.2cm}
\end{figure}

A similar situation occurs for the 3rd order cumulant $C_3$, as
illustrated in Fig. \ref{fig:frcC3}. We thus conclude that the strong
suppression of the effects of fluctuations that we observe in our results
cannot be imputed to the fact that the coupling is effectively weaker in
the RC scenario than in the FC one (for a same value of $\alpha$ at
$Y=0$).

Incidentally, since the $Y$--dependence of our results for $\sigma^2(Y)$
appear to be rather well described, at least at large $Y$, by the
expression in Eq.~\eqref{eq-sigmarun}, it is interesting to perform a fit
based on this expression and thus extract the value of the front
diffusion coefficient from the numerical results. We have repeated this
fit for several values of $\beta$, for rapidities within the range $200 <
Y/\beta < 300$, with the results shown in the right hand side of Fig.
\ref{fig:frcsigma}. As indicated there, these fits suggest the scaling
law $D(\beta)\propto 1/\beta$, for which we have no fundamental
understanding.

\vspace*{.3cm}
\texttt{(ii)} {\em The average front position and velocity}
\vspace*{.2cm}

\begin{figure}[t]
    \centerline{
    \includegraphics[width=7.4cm,angle=-90]{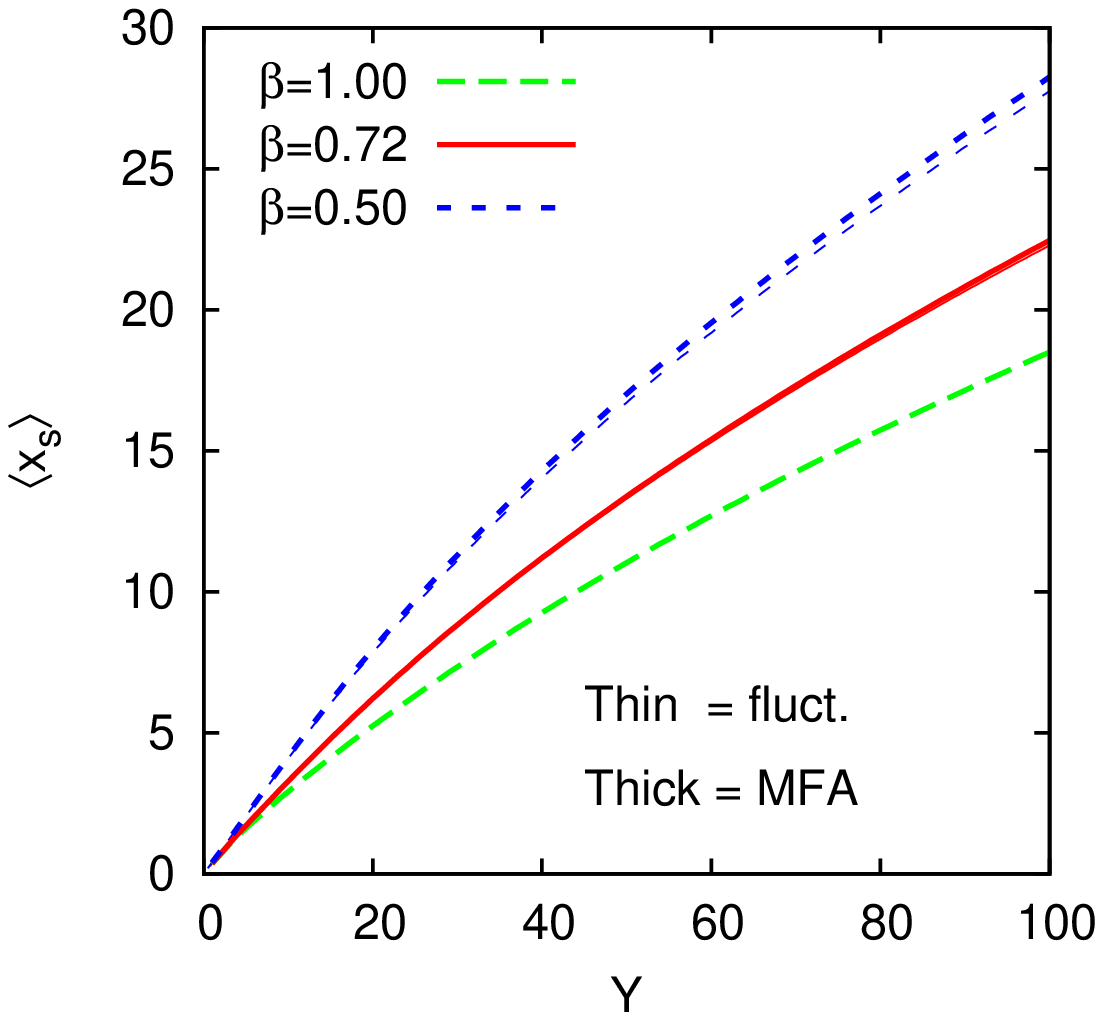}
    \includegraphics[width=7.4cm,angle=-90]{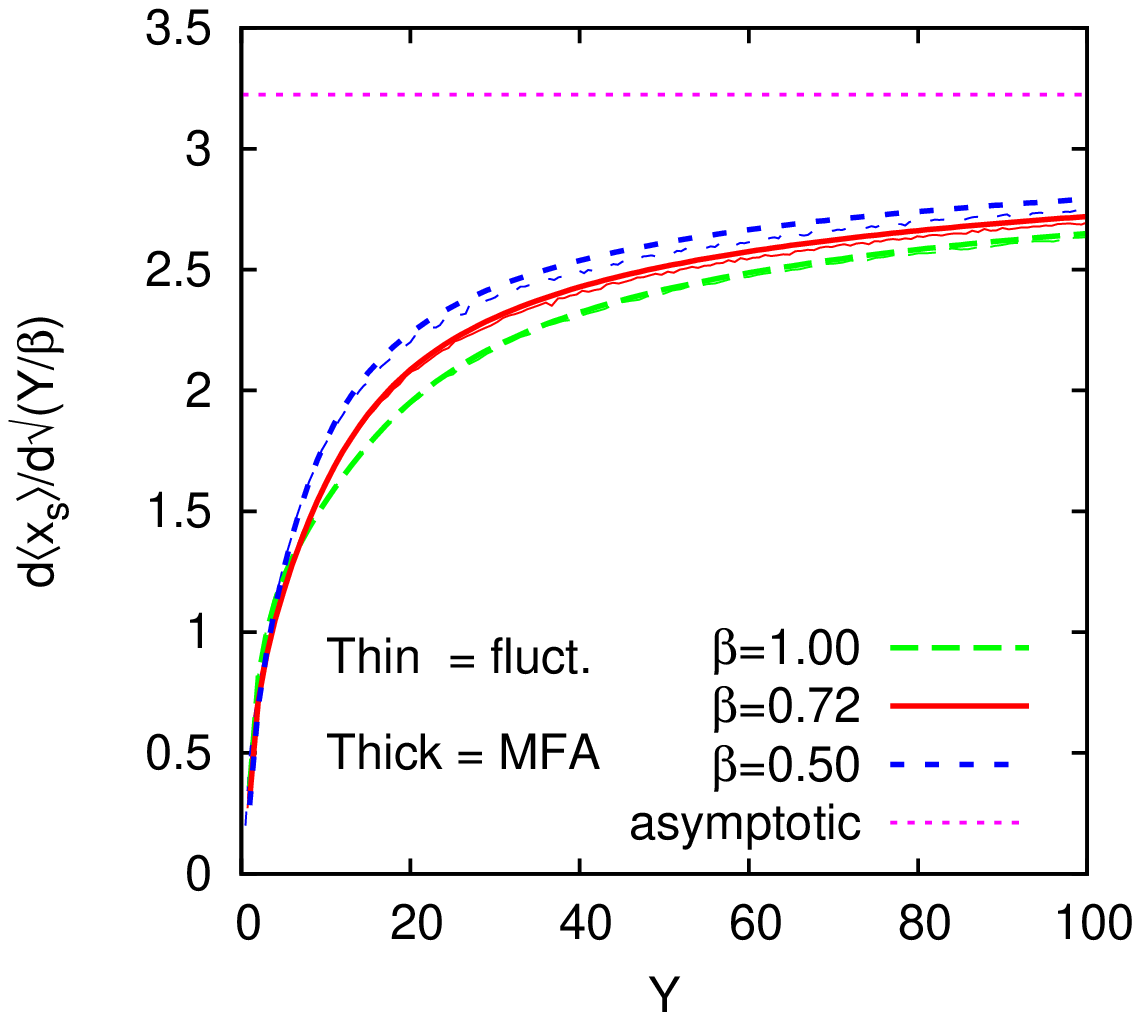}}
    \caption{\small
    The average front position (left) and the corresponding velocity
    (right) for the running coupling evolution: the results of the full
    evolution, including fluctuations (thin lines) are compared to the
    respective mean--field results (thick lines).
    }\label{fig:rcxs}
\end{figure}

Further evidence for the suppression of fluctuations comes from a study
of those quantities which exist already at mean field level, like the
(average) position of the front $\avg{x_s}$ and the corresponding
velocity. Our respective numerical results are displayed in Fig.
\ref{fig:rcxs}, for both the complete evolution and its mean field
approximation. These results should be compared to the corresponding ones
at FC, in Fig.~\ref{fig:fcxs}. In the latter, the difference from the
mean field behavior is manifest, and is increasing with $Y$. In  Fig.
\ref{fig:rcxs}, on the other hand, there are only minor differences
between the full evolution and the MFA: the respective curves fall almost
on top of each other, for all values of $Y$. (The small visible
differences are in fact consistent with the magnitude of the dispersion
reported in Fig. \ref{fig:rcsigma}.) In particular, on the right hand
figure, one sees that all the velocity curves remain well below the
limiting value $v_0=\sqrt{2\lambda_0}\approx 3.22$ which is expected at
asymptotically large $Y$ both in the MFA, cf. Eq.~\eqref{eq-vrc}, and in
the full evolution with fluctuations. These results have the following
implications:

\texttt{(a)}  For the considered range in rapidities, both types of
evolution (with or without fluctuations) are still far away from the
asymptotic regime.

\texttt{(b)} In the full evolution including fluctuations, the rapidities
covered by our simulations are too small for the fronts to be fully
formed. (Since, if the fronts were fully formed at some intermediate
value of $Y$, then their subsequent evolution would have been quite
different from the mean field case.) In particular, the fact that the
measured front velocity remains significantly smaller than the asymptotic
value $v_0$ is not to be attributed to `cut--off' effects due to
fluctuations, but rather to preasymptotic effects, which for $Y< Y_{\rm
form}$ are the same with or without fluctuations.


Conclusion \texttt{(b)} above can be reformulated by saying that the
front `formation time' $Y_{\rm form}$ should be larger than, or
marginally comparable to, the maximal rapidity which is covered by our
analysis. This is also consistent with the previous results concerning
the smallness of fluctuations: as explained in Sect. \ref{sect-analytic},
the fluctuations cannot significantly develop until the individual fronts
have been fully formed.

The fact that the front `formation time' $Y_{\rm form}$ can be relatively
large at RC should not be a surprise, in view of the discussion in Sect.
\ref{sect-analytic}. What is quite surprising, however, is that this
quantity can be {\em that} large, namely of $\order{100}$ or even larger.
(In fact, in our simulations, we have never seen any sign that the fronts
have been formed, until the highest rapidities.) To further consolidate
this discovery, we performed another numerical test, which directly
measures the front region, and thus the formation time:

\vspace*{.3cm}

\texttt{(iii)} {\em The reduced front: event--by--event geometric
scaling} \vspace*{.2cm}

By inspection of the analytic results in Sect. \ref{sect-analytic} (see,
especially, Eqs.~(\ref{eq-satsol}) and (\ref{eq-tsolmft}) there), one
sees that a study of the {\em reduced front} $T_{\rm red}(x,Y)\equiv
\rme^{\gamma_s (x-x_s)}T(x,Y)$ can give us access to the width of the
geometric scaling window, for both fixed and running coupling. Indeed,
for $x\simle x_s$, which includes the saturation region and the
transition region around $x\sim x_s$, where the shape of the amplitude is
not analytically known, $T_{\rm red}(x,Y)$ vanishes exponentially,
whereas for very large $x$, such that $x-x_s \gg x_{\rm diff}$, it
rapidly vanishes once again, due to diffusion. On the other hand, within
the window for geometric scaling, $T_{\rm red}\propto (x-x_s + c)$ grows
linearly with $x-x_s$, cf. Eq.~\eqref{eq-tscale}; hence, the width of the
geometric scaling window can be identified as the region of linear
increase for $T_{\rm red}(x,Y)$.

\begin{figure}[t]
    \centerline{
    \includegraphics[width=7.4cm,angle=-90]{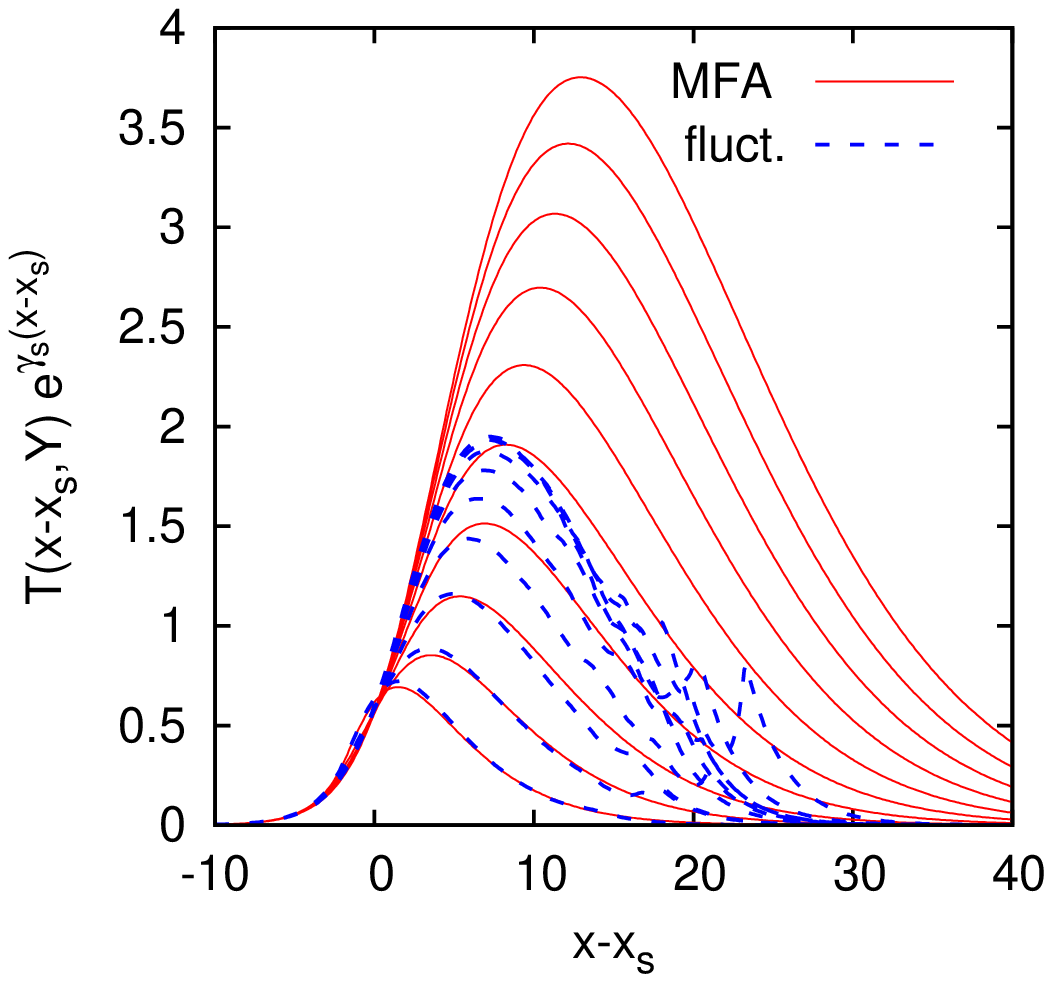}
    \includegraphics[width=7.4cm,angle=-90]{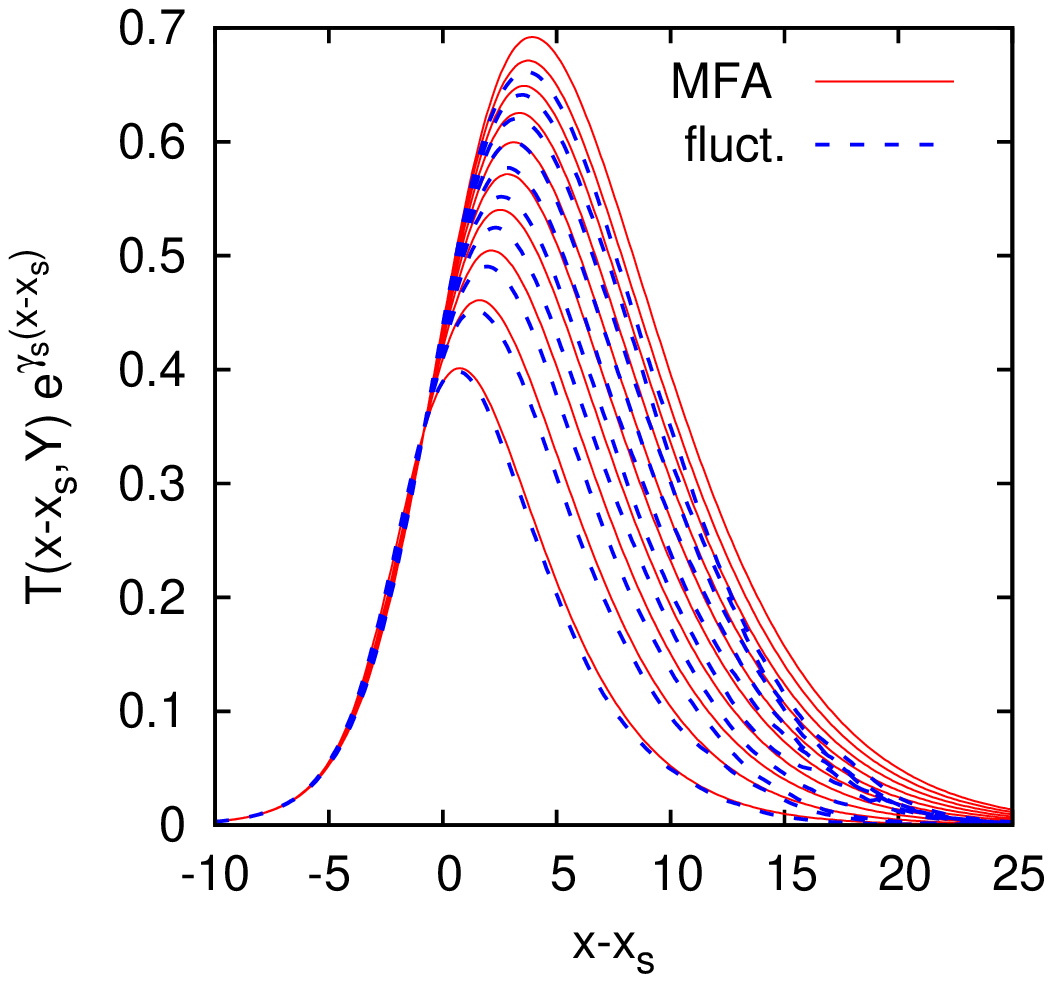}}
    \caption{\small
    The reduced front as a function $x-x_s$ for different values of $Y$,
    with of without fluctuations (in the fluctuation case,
    we show the respective average quantity).
    Left: FC evolution with $\alpha=0.1$
    (from left to right: $Y=5,10,\dots\,50$).
    Right: RC evolution with $\beta=0.72$
    (from left to right: $Y=10,20,\dots\,100$).
    Notice the different scales used in the two figures.
    }\label{fig:reduced}
\end{figure}

We have measured the reduced fronts with both FC and RC, with the results
shown in Fig. \ref{fig:reduced}. For fixed coupling (figure on the left),
there is a clear difference between the MFA and the complete evolution
including fluctuations: whereas in the former case, the scaling window
keeps growing with $Y$, and quite fast (recall that the respective
diffusive radius is expected to grow like $\sqrt{Y}$, cf.
Eq.~\eqref{eq-winscalefc}), in the presence of fluctuations, the growth
of the scaling region is similar in the early stages, but then it
saturates, around $Y\sim 20$, at a maximal value which defines the width
of the front. The fluctuations in the tail of the front at large $x-x_s$
are clearly visible.

Moving to the RC case (right figure), the situations looks very
different: the curves corresponding to mean--field and, respectively,
fluctuations are now very close to each other, no fluctuations are
visible in the tail, and, besides, the width of the scaling region is
much smaller than for FC, as expected (since with RC this region rises
only like $Y^{1/6}$, cf. Eq.~\eqref{eq-winscale}). The fact that, in the
fluctuation case, the width of the scaling region keeps growing with $Y$
confirms that the front has not been formed, in agreement with the
previous results in this section. In fact, the maximal extent of the
geometric scaling window as seen on this plot is about $x-x_s\simeq 5$
for $Y=100$, which is rather small --- at most, marginally comparable to
our previous estimate, Eq.~\eqref{eq-LRC}, for the front width $L$
evaluated for $\alpha=0.1$.

 \vspace*{.3cm}

\texttt{(iv)} {\em Approximate geometric scaling for the average
amplitude} \vspace*{.2cm}

\begin{figure}[t]
    \centerline{
    \includegraphics[width=14.cm]{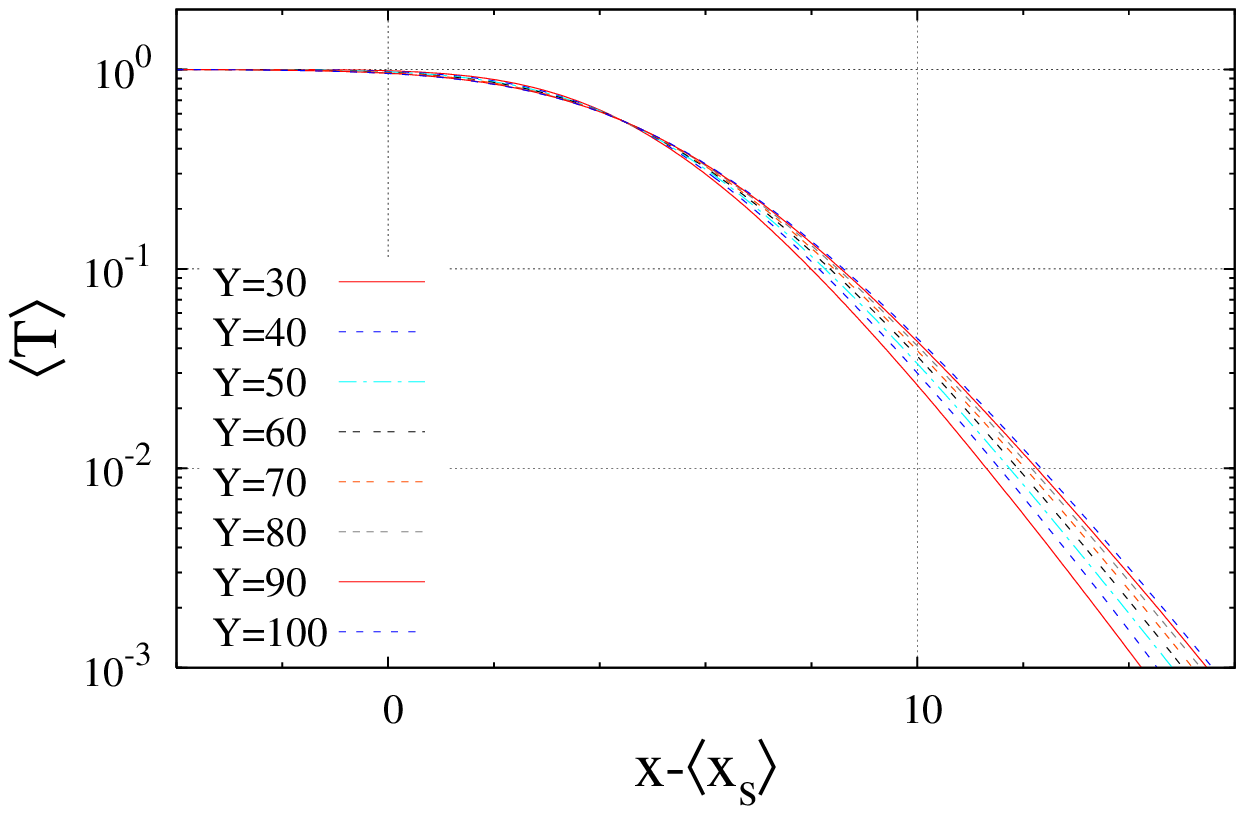}}
    \caption{\small
    The average amplitude for the RC evolution with $\beta=0.72$
    represented as a function of the scaling variable $x-\avg{x_s}$,
    for various values of $Y$.
    }\label{fig:geom}
\vspace*{.2cm}
\end{figure}

Since our numerical results show very small dispersion, cf. Fig.
\ref{fig:rcsigma}, it is quite clear that the shape of the average
amplitude $\avg{T(x)}_Y$ remains close to that of any of the individual
fronts which compose the ensemble, and hence to the shape of the
mean--field front. We thus expect $\avg{T(x)}_Y$ to exhibit {\em
approximate geometric scaling}, according to the pattern described below
Eq.~(\ref{eq-tsolmft}). In order to check this, we have displayed in Fig.
\ref{fig:geom} the average amplitude emerging from our RC simulations
(with $\beta=0.72$) as a function of the scaling variable $z\equiv
x-\avg{x_s}$, for various values of $Y$ up to $Y_{\rm max}=100$. From
this figure, one sees that the various curves fall almost on top of each
other for $z\simle 5$, but they start to deviate from each other for
larger values of $z$, although this deviation remains quite small. This
behavior is indeed consistent with both the analytic considerations in
Sect. \ref{sect-analytic} and the previous numerical results in this
present section. Namely, the value $z\simeq 5$ for the upper bound of the
geometric scaling window is in agreement with our former results for the
reduced front, as shown in the right plot in Fig. \ref{fig:reduced}.
Also, the fact that the dispersion between the various curves (at $z >
5$) is decreasing when increasing $Y$, as it can be checked via a close
inspection of Fig. \ref{fig:geom}, is in agreement with our theoretical
expectation that the violation of geometric scaling proceeds via BFKL
diffusion (cf. Eq.~(\ref{eq-tsolmft})).

The quality of the geometric scaling behavior\footnote{This quality could
be more precisely characterized with the method proposed in Ref.
\cite{GPSS06}.} visible in Fig. \ref{fig:geom} is considerably lower than
for the corresponding {\em mean--field} results at FC, but much better
than it would be for the {\em full} FC evolution, including fluctuations.
(For $Y=100$, the stochastic FC evolution with $\alpha=0.1$ would
generate a dispersion $\sigma^2\sim 10$, which is large enough for the
geometric scaling to be completely washed out in the average amplitude,
and replaced by diffusive scaling; cf. Fig.~\ref{fig:fcsigma} and the
discussion in Ref. \cite{ISST07}.) In fact, the amount of geometric
scaling violation in Fig. \ref{fig:geom} is indeed comparable to the one
observed in the small--$x$ data at HERA \cite{geometric,MS06,GPSS06}.

We conclude this section with a lesson from our analysis which may shed
some light on a rather surprising recent finding in Ref. \cite{Alb07}:
namely, in that work, an unexpectedly large `anomalous dimension' was
reported, $\gamma\simeq 0.85$, based on an exponential fit to the fronts
generated by the BK equation with running coupling. This value
$\gamma\simeq 0.85$ is significantly larger than the saturation anomalous
dimension $\gamma_s\simeq 0.63$ which is expected to control the decay of
the amplitude in the geometric scaling region. Or, as we have seen in
both Fig. \ref{fig:reduced} (right figure) and Fig. \ref{fig:geom}, the
window for true scaling is in fact quite narrow: $z\simle 5$ at $Y=100$;
outside this window, the amplitude decays much faster, because of the
diffusion term. It is therefore likely that the result reported in Ref.
\cite{Alb07} is the consequence of enforcing an exponential fit over a
relatively large window in $z$, much larger than the actual window for
geometric scaling.

\newpage
\section{Conclusions and outlook}\label{sect-conc}
\setcounter{equation}{0}

In this paper we have presented the first analysis of the consequences of
the running of the coupling on the high--energy evolution with Pomeron
loops, in the context of a simple one--dimensional model which captures
the relevant dynamics in QCD.  We have found that the Pomeron loop
effects are strongly suppressed by the running of the coupling, up to the
highest energies that we have investigated --- which go well beyond the
energies of interest for the phenomenology of QCD. The main reason for
this surprising behavior is the fact that, in the presence of a running
coupling, the diffusive radius grows very slowly with rapidity.
Therefore, the wave fronts preserve a pre--asymptotic shape which is not
favorable for the growth of fluctuations. During this pre--asymptotic
evolution, which in our simulations extends up to $Y \simeq 200$, the
dynamics is similar to the respective prediction of the mean--field
approximation (with running coupling, of course). In particular, due to
the suppressed dispersion, approximate geometric scaling is preserved for
the average scattering amplitude. The window for strict geometric scaling
is rather narrow, but an approximate scaling behavior extends well
outside this window. We believe that this should suffice to explain the
respective scaling observed in the HERA data, but a firm conclusion in
that sense would deserve a dedicated study.

In the course of our analysis, we have performed various numerical tests
to verify that our conclusions are insensitive to the details of the
numerical procedure (e.g., the discretization of the spatial axis), to
the prescription used the `freeze' the running of the coupling in the
infrared, and also to the specific implementation of a running coupling
in the context of the model (in addition to our main choice,
Eq.~\eqref{eq-tau}, for the running of $\alpha^2$, we have also performed
calculations with some other prescriptions, and obtained similar
results).

It is also useful at this stage to summarize the limitations of our
present analysis, and thus pave the way towards further studies: This
study has been restricted to asymmetric initial conditions (dilute
projectile vs. dense target), where the projectile corresponds to a
dipole in QCD. This setup covers those physical problems which, in QCD,
admit a dipolar factorization --- chiefly among them, deep inelastic
scattering, but also particle production at forward rapidities in
hadron--hadron ($pp$, $pA$) collisions. However, our current model can
accommodate arbitrary initial conditions, including more symmetric ones,
and it would be interesting to study the influence of the initial
conditions on our main conclusion (the suppression of Pomeron loop
effects by the running of the coupling). This could shed more light on
the physical mechanism responsible for this suppression and, in
particular, consolidate the argument about the front formation time that
we have proposed, and which seems to be supported by the present
analysis.

Furthermore, it would be important to clarify the model--dependence of
our conclusions, if any. If the extension of the present analysis to real
QCD (which would require solving the Pomeron loop equations of Refs.
\cite{IT04,MSW05}) looks still prohibitive for the time being, it would
be nevertheless interesting to apply a similar study to other
one--dimensional models of the `reaction--diffusion' type, and thus check
whether the universality expected for such models still persists after
the inclusion of running coupling effects. This may have physical
consequences for other kinds of problems, say in the framework of
statistical physics, where what we call `running coupling' in the context
of QCD would actually correspond to an inhomogeneous medium, in which the
rates for particle splitting and merging scale with $x$ like (powers of)
$1/x$.

\section*{Acknowledgments}

We would like to thank Al Mueller for fruitful discussions and insightful
remarks on our early numerical results. We acknowledge useful discussions
with Guillaume Beuf and Robi Peschanski. L.P. would like to acknowledge
CAPES for financial support. G.S. is funded by the National Funds for
Scientific Research (FRNS, Belgium). This manuscript has been authored
under Contract No. DE-AC02-98CH10886 with U.S. Department of Energy.


\end{document}